\documentclass[useAMS,usenatbib,referee,letterpaper]{mn2e}
\pdfoutput=1
\usepackage{aas_macros}
\usepackage{graphicx}
\usepackage{deluxetable}
\usepackage{url}
\usepackage[toc,page]{appendix}
\title[{Very hot H-deficient [WCE]-type CSPNe}]
{UV spectral analysis of very hot H-deficient [WCE]-type central stars of planetary 
nebulae: NGC 2867, NGC 5189, NGC 6905, Pb 6, and Sand 3}

\author[Keller et al.]
{Graziela R. Keller$^{1,2}$\thanks{E-mail:
graziela.keller@iag.usp.br, visiting scientist at JHU Department of Physics and 
Astronomy.}, Luciana Bianchi$^2$ and 
Walter J. Maciel$^1$ \\
$^1$Instituto de Astronomia,
Geof\'{\i}sica e Ci\^{e}ncias Atmosf\'{e}ricas, Universidade de S\~{a}o Paulo, 
Cidade Universit\'{a}ria, S\~{a}o Paulo/SP, Brazil.\\
$^2$Department of Physics and
Astronomy, The Johns Hopkins University, 3400 N. Charles Street, Baltimore, MD 
21218, USA.}

\date{Released 2014 Xxxxx XX}

\pagerange{\pageref{firstpage}--\pageref{lastpage}} \pubyear{2014}

\begin{document}
\label{firstpage}
\maketitle

\begin{abstract}
We analysed UV FUSE, IUE, and HST/STIS spectra of five of the hottest [WCE]-type 
central stars of planetary nebulae: NGC 2867, NGC 5189, NGC 6905, Pb 6, and Sand 3. The analysis leveraged on our grid of CMFGEN synthetic spectra, 
which covers the parameter regime of hydrogen deficient central stars of 
planetary nebulae and allows a uniform and systematic study of the stellar 
spectra. The stellar atmosphere models calculated by us include many elements 
and ionic species neglected in previous analyses, which allowed us to improve 
the fits to the observed spectra considerably and provided an additional 
diagnostic line: the Ne\,{\sc vii} $\lambda$ 973 $\mathrm{\AA}$, which had not been 
modelled in [WCE] spectra and which presents, in these stars, a strong P-Cygni 
profile.  

We report newly derived photospheric and wind parameters and elemental 
abundances. The central stars of NGC 2867, NGC 5189, and Pb 6 had their 
temperatures revised upward in comparison with previous investigations and we 
found the carbon to helium mass ratio of the sample objects to span a wide range 
of values, $0.42<C:He<1.96$. Modelling of the Ne\,{\sc vii} $\lambda$ 973 
$\mathrm{\AA}$ P-Cygni profile indicated strong neon overabundances for the 
central stars of NGC 2867, NGC 5189, NGC 6905, and Pb 6, with Ne mass 
fractions between 0.01 and 0.04. Nitrogen abundances derived by us for the 
central stars of NGC 5189, Pb 6, and Sand 3 are higher than previous 
determinations by factors of 3, 10, and 14, respectively.         
\end{abstract}


\section{Introduction}

Central stars of planetary nebulae, or CSPNe, are evolved stages of low- and 
intermediate-mass stars. They are very 
hot, with surface temperatures between 20,000 and 200,000 K, and many of them 
present signatures of strong stellar winds in their spectra. It is generally 
believed that the mass loss observed in these stars is driven by radiation 
pressure in spectral lines, the same mechanism thought to accelerate the winds 
of hot massive stars and which is intrinsically unstable and leads to the 
formation of density inhomogeneities usually called clumps \citep[see, for 
example,][]{2008cihw.conf...17M,2002A&A...383.1113D,2003A&A...406L...1D,
2005A&A...437..657D,2007A&A...476.1331O,Sundqvist2013}.

CSPNe are the connection between the asymptotic giant branch (AGB) phase and the 
white dwarfs (WDs) and are a unique opportunity to test stellar atmosphere, 
structure and evolution models. Their winds have the potential to impact 
subsequent stellar evolution and shape the planetary nebulae, which also owe 
their ionization to the strong radiation fields of the central stars. Thus, 
solid determinations of the photospheric
and wind parameters of CSPNe, as well as of their chemical abundances, can help 
answering questions concerning stellar
evolution, in particular the connection among the different types of hydrogen 
deficient central stars, the wind driving mechanism, and the surrounding 
nebulae.

It is estimated that about 30 per cent of CSPNe are hydrogen deficient 
\citep{Weidmann2011}. The main constituents of their atmospheres are, most 
commonly, helium, carbon, oxygen, neon, and, often, also nitrogen. Among the 
hydrogen deficient CSPNe, those evolving away from the AGB towards higher 
temperatures (in the Hertzsprung-Russell diagram - HRD), showing strong carbon 
and helium emission lines in their spectra, similar to massive Wolf-Rayet stars, 
are named [WC] stars, which are further divided into early ([WCE]) and late 
types ([WCL]), according to the ionization stages visible in their spectra. The 
hydrogen deficient CSPNe positioned at the top of the white dwarf cooling track 
in the HRD, exhibiting weak, or no wind features, in addition to absorption 
lines of highly ionized helium, carbon, and oxygen, belong to the 
PG1159 class. Due to the proximity of these types of 
hydrogen deficient CSPNe in the $\log T_{\ast}-\log g$ diagram and also because 
of similarities in their abundance patterns, they are thought to form an 
evolutionary sequence, in which they would move away from the AGB and progress as [WC] 
stars towards higher temperatures, evolving from [WCL] to [WCE] types as their 
temperatures increase, until nuclear burning ceases and they enter the white 
dwarf cooling track. Their luminosities and mass-loss rates decrease and their 
winds reach the very high terminal velocities (up to 4000 km s$^{-1}$) typical 
of [WC]-PG1159 and PG1159 stars 
\citep{1991A&A...247..476W,2000A&A...362.1008G,2001A&A...367..983P}. Other 
classes of H-deficient post-AGB objects exist, which will not be discussed here 
\citep[see, for 
example,][]{2007AJ....134.1380M,2010AIPC.1273..219T,2013MNRAS.430.2302T,
1998AeA...338..651R}.

He, C, O, and Ne abundances observed in [WC] and PG1159 stars - which, however, 
vary greatly from one object to the other - resemble those of the intershell 
region of AGB models. According to \citet{Werner2007a}, in the case of 
PG1159 stars, the observed abundance intervals are: $X_{He}=0.30-0.85$, 
$X_{C}=0.15-0.60$, and $X_{O}=0.02-0.20$. Abundance determinations for [WC] stars 
are approximately in the intervals: $X_{He}=0.40-0.79$, $X_{C}=0.15-0.55$, and 
$X_{O}=0.01-0.15$ 
\citep{1996A&A...312..167L,1997IAUS..180..114K,1997A&A...320...91K,
1998A&A...330..265L,1998A&A...330.1041K,2007ApJ...654.1068M}. Post-AGB
evolutionary models are able to predict the helium, carbon, and oxygen abundance 
patterns
seen in PG1159 and [WC] stars, explaining the wide range of values seen by 
differences in the
initial mass of the stars and on the number of thermal pulses suffered during 
the AGB phase
\citep[see, for example,][and references therein]{Werner2007a}. Their hydrogen 
deficiency is usually thought to be the result of late thermal pulses, which can 
occur either at the tip of the AGB (in which case it is named an AGB final 
thermal pulse or AFTP), or during the post-AGB evolutionary phase, as in the 
very late thermal pulses (VLTP) of the born-again scenario of 
\citet{1983ApJ...264..605I} (taking place after the star has entered the WD 
cooling track in the HRD), or in the so called late thermal pulses (LTP) 
occurring during the CSPN phase \citep[see][]{2001Ap&SS.275....1B, 
2001Ap&SS.275...15H, 2005A&A...435..631A}. Each of these scenarios produces 
H-deficient stars. However, the surface abundance patterns originated will 
differ between the scenarios \citep[see, for example,][and references 
therein]{2006PASP..118..183W}.

\begin{figure}
\centering
\includegraphics[scale=0.48]{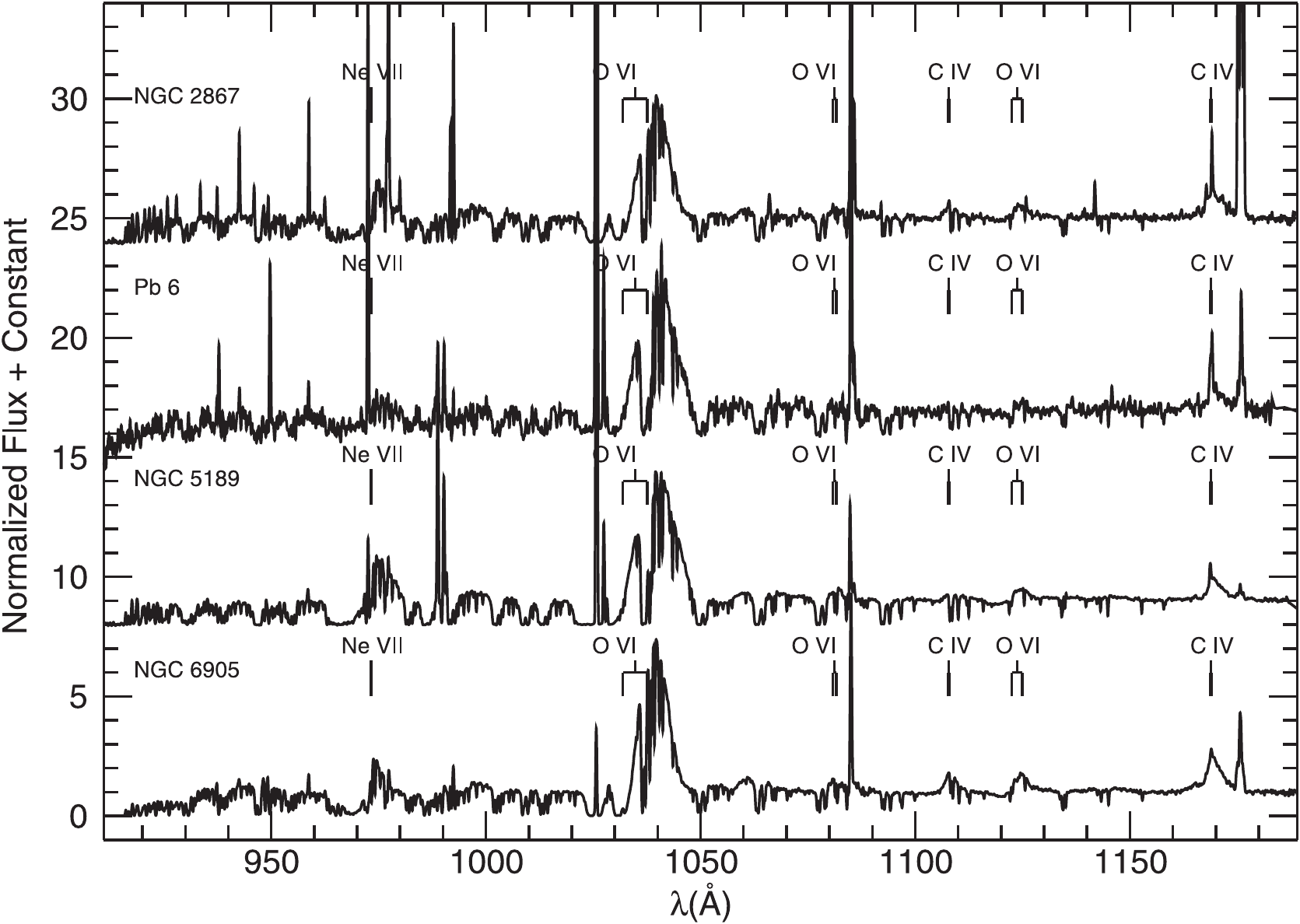}
\includegraphics[scale=0.48]{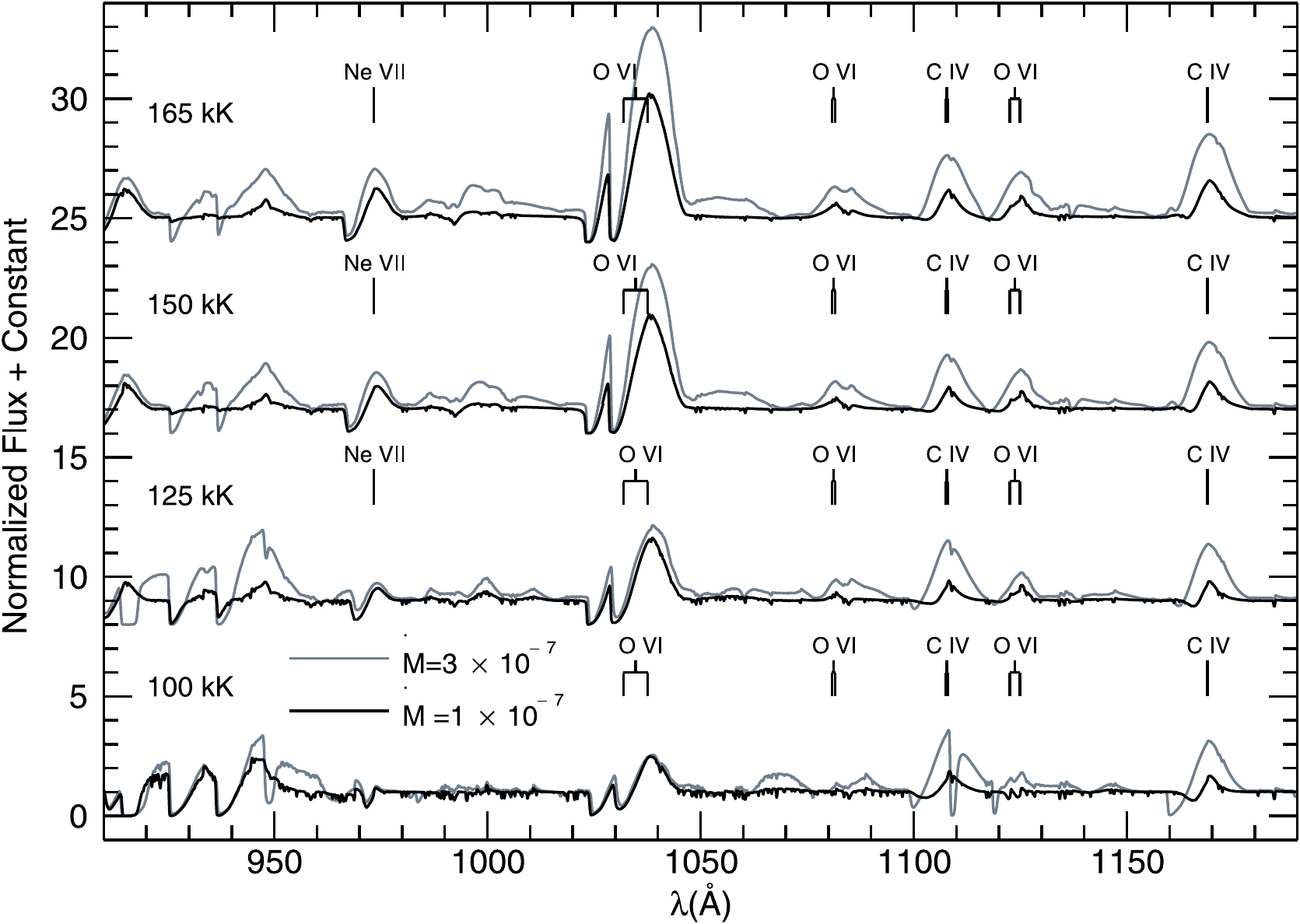}
\caption{Left panel: FUSE spectra of the [WCE] CSPNe NGC 6905, NGC 5189, Pb 6 
and NGC 2867. The numerous narrow absorptions are due to interstellar H$_{2}$, 
which affects the blue edge of the Ne\,{\sc vii} P-Cygni profile. Right panel: sample 
synthetic spectra from our model grid, calculated for different values of 
stellar temperature and two different mass-loss rates (in units of M$_{\odot}$ 
yr$^{-1}$). All models shown adopt $v_{\infty}=2500$ km s$^{-1}$ and models of 
the same temperature adopt the same stellar radius. The observed and synthetic 
spectra shown were convolved with a Gaussian of 0.2 $\mathrm{\AA}$ FWHM for 
clarity.}\label{FUSEspectra}
\end{figure}

\begin{figure}
\centering
\includegraphics[width=0.48\columnwidth]{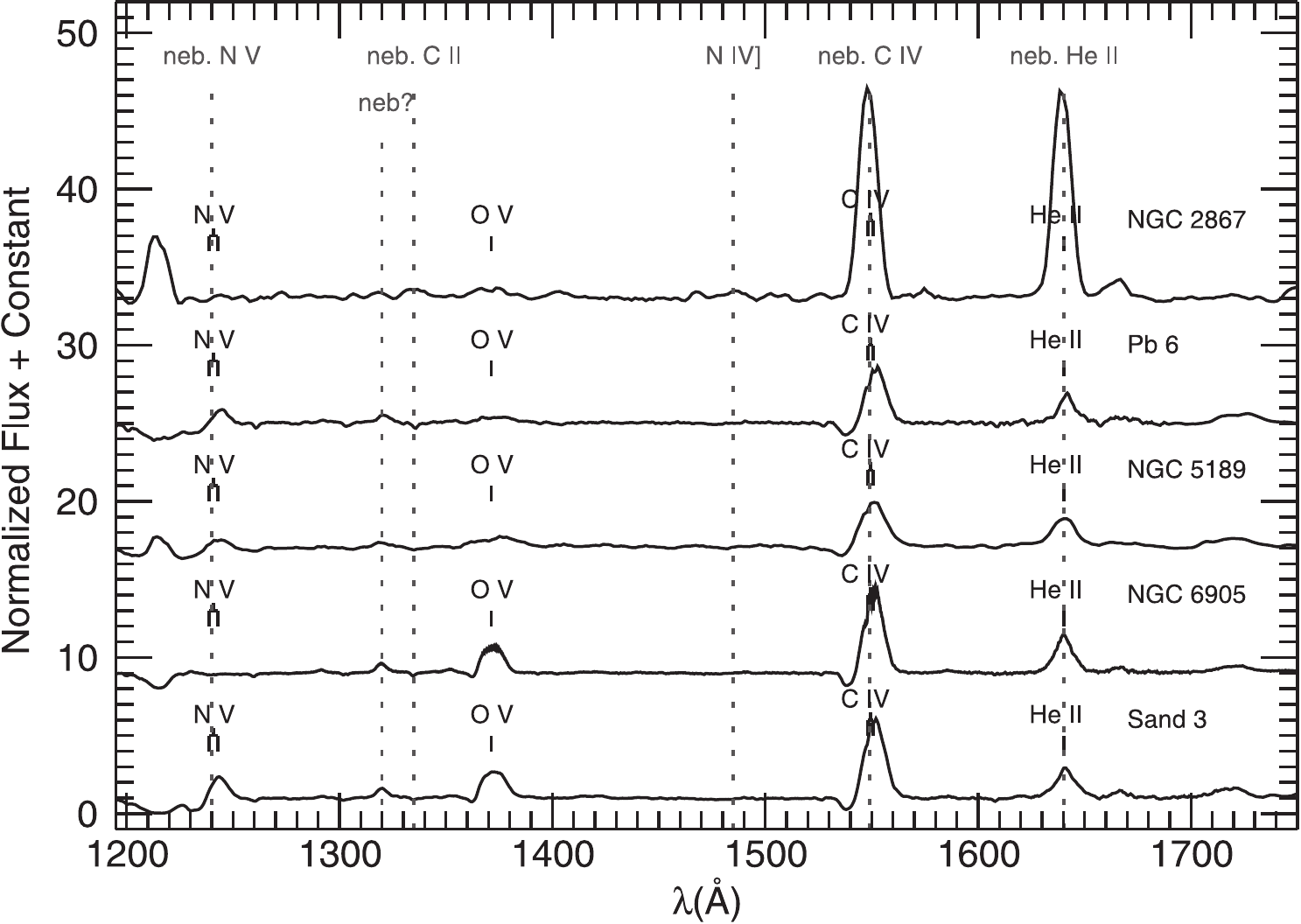}
\includegraphics[width=0.48\columnwidth]{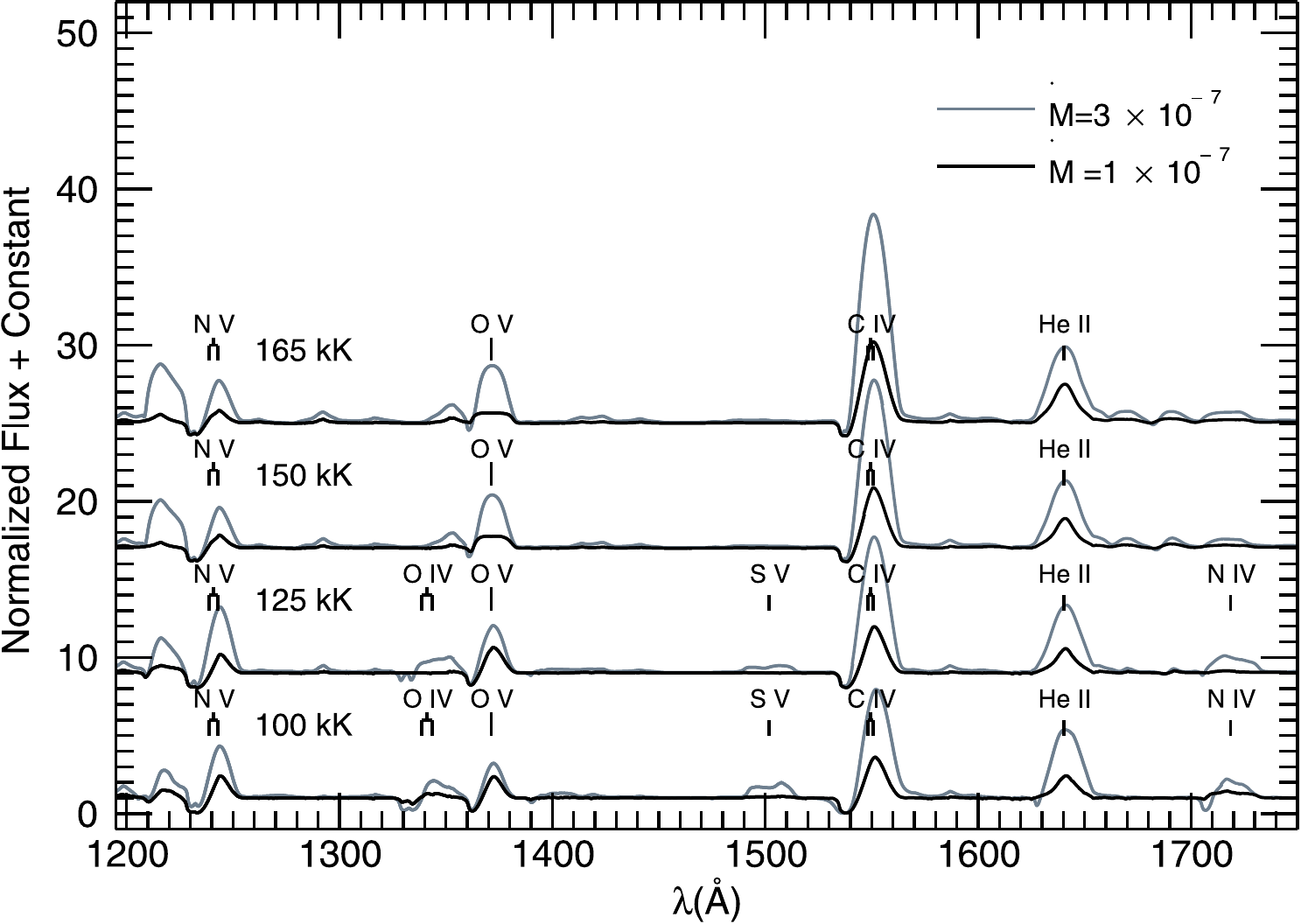}
\caption{Left panel: HST/STIS G140L spectra of the [WCE] CSPNe NGC 6905, Pb 6 
and Sand 3, and low resolution IUE spectra of NGC 5189 and NGC 2867. The dotted 
vertical lines indicate the location of emission lines commonly seen in PNe. 
\citet{Feibelman1996} identified the structure around 1320 $\mathrm{\AA}$ as 
being a nebular [Mg V] line in the spectra of Sand 3. We, however, find evidence 
of its stellar origin, as we discuss on Section \ref{sec:targets}. Right panel: 
sample synthetic spectra from our model grid (same models as in Fig. 
\ref{FUSEspectra}), between 1200 and 1750 $\mathrm{\AA}$, calculated for 
different values of stellar temperature and two different mass-loss rates (in 
units of M$_{\odot}$ yr$^{-1}$). The synthetic spectra shown were convolved with 
a HST/STIS G140L instrumental LSF.}\label{UVspectra}
\end{figure}

[WCE] CSPNe are thus the hottest among [WC]-type stars and are probably the 
direct predecessors of [WC]-PG1159 and PG1159 stars, which are, in turn, 
believed to evolve into DA and DO WDs \citep{Althaus2010}. Their 
analysis can help establish constraints for the different post-AGB evolutionary 
scenarios. The spectra shown by the hottest [WCE] stars are characterized by 
lack of photospheric absorption lines, the presence of strong, broad emission 
lines, UV P-Cygni profiles, and a paucity of lines in the optical region, in 
which none of the few strong lines they present have P-Cygni profiles. 
Essential diagnostic lines lay shortwards of Ly\,$\alpha$, in the far-UV range, 
which was provided by FUSE (Far Ultraviolet Spectroscopic Explorer - 905$-$1187 
$\mathrm{\AA}$ - \citet{2000ApJ...538L...1M}). In this region, these stars show 
two conspicuous P-Cygni profiles: Ne\,{\sc vii} $\lambda$ $973.3$ 
$\mathrm{\AA}$ and O\,{\sc vi} $\lambda \lambda$ 1031.9, 1037.6 $\mathrm{\AA}$, as shown in Fig. \ref{FUSEspectra}. Complementary line diagnostics 
are found in the spectral region between 1150 and 3200 $\mathrm{\AA}$, covered 
by IUE (International Ultraviolet Explorer) and HST (Hubble Space Telescope) 
spectrographs (Fig. \ref{UVspectra}). A detailed description of the UV spectra 
of [WCE] CSPNe was provided by \citet{kelleretal2011PORT}.

In this paper, we present a comparative analysis of UV and far-UV spectra of the 
hot [WCE]-type CSPNe NGC 2867, NGC 5189, NGC 6905, Pb 6, and Sand 3 (WO1 stars in the unified classification 
scheme of \citet{1998MNRAS.296..367C}), prompted by the lack of a comprehensive, systematic analysis of [WCE] 
CSPNe making use of the available high resolution spectra and of today's modern 
stellar atmosphere codes. Our analysis includes many ionic species neglected in 
previous works, thus allowing us to model, for the first time in [WCE] stars, 
the Ne far-UV spectral lines, besides improving the overall match between model 
and observations. The analysis was based on our extensive grid of synthetic 
spectra \citep{kelleretal2011PORT} and numerous subsequent model 
calculations for each individual object. A detailed analysis of 
the central star of NGC 6905 was presented in Keller et al. (2011) and the results are now compared with those obtained for the other stars studied in this work, 
placing the central star of NGC 6905 in the context of [WCE] 
CSPNe. With the exception of Pb 6, all CSPNe studied here are classified as GW 
Vir (=PG 1159$-$035) variables - a class of nonradial pulsating 
variable H-deficient pre-White Dwarfs, showing variability $\la$0.3 mag \citep[see][and references therein]{Althaus2010}.

The paper is organized as follows: in Section \ref{sec:targets}, we describe the 
sample objects and the data used in the analysis. Section \ref{sec:thegrid} 
describes the synthetic spectra. In Section \ref{sec:SpecAnalyses}, we 
describe the spectral analysis of the sample objects. Finally, we present our 
conclusions in Section \ref{sec:conclusions}.

\section{Sample objects and data}
\label{sec:targets}

The objects studied here, the central stars of NGC 2867, NGC 5189, NGC 6905, Pb 6, and Sand 3, belong to the small group of [WCE] stars for which good 
quality UV spectra exist: quality far-UV FUSE spectra exist for four of these 
stars (NGC 2867, NGC 5189, NGC 6905, and Pb 6). However, only for NGC 6905 and Pb 
6 there are complementary HST/STIS UV spectra. In the case of NGC 5189 and NGC 
2867, we used complementary IUE low resolution spectra. Sand 3, on the other 
hand, has excellent HST/STIS G140L and G230MB spectra, but no FUSE, which limits 
the available line diagnostics, but its analysis is nonetheless interesting, 
particularly for its spectral similarity to the other stars, especially to the 
central star of NGC 6905. The available STIS spectra of Sand 3 were complemented 
by low resolution IUE spectra between 1725 and 3300 $\mathrm{\AA}$.

Our sample objects are listed in Table \ref{objects}, along with radial 
velocities ($v_{rad}$), distances, and nebular parameters (expansion velocity - 
$v_{exp}$, angular diameter, and shape). Throughout this paper, the observed 
spectra presented in the figures have been shifted to the rest frame of the 
star, using the listed radial velocities. The utilized spectra are listed in 
Table \ref{spectra}. IUE, HST/STIS, and FUSE spectra used in the analysis were retrieved from  
MAST (Mikulski Archive for Space Telescopes - \url{http://archive.stsci.edu/}). NGC 2867, NGC 5189, and Pb 6 FUSE and Pb 6 and Sand 3 STIS spectra have not been 
used in stellar atmosphere modelling before this work. NGC 6905's STIS spectra, 
as well as the region of its FUSE spectra shortwards of 1000 $\mathrm{\AA}$, in which 
the strong P-Cygni profile of the $\lambda$ $973.3$ $\mathrm{\AA}$ Ne\,{\sc vii} line 
can be found, were first modelled by us and presented in 
\citet{kelleretal2011PORT}. The objects were thus selected among the known 
sample of hot [WCE] CSPNe for the availability of good UV spectra, most of them 
not previously analysed.

\begin{table}
\caption{Sample objects.}
\label{objects}
\begin{tabular}{llllll}
\hline
Object & v$_{rad}$ [km s$^-1$]  & Dist. [kpc] & Nebular Diam. [arcsec] & 
v$_{exp}$ [km s$^-1$] & Shape \\
\hline
NGC 6905 &  -8.4\tablenotemark{a} & 1.73\tablenotemark{e} ; 
1.75\tablenotemark{f} ; 1.80\tablenotemark{g} & 43.3 $\times$ 
35.6\tablenotemark{d} & 43.5\tablenotemark{a} & Elliptical \\
NGC 5189 &  -9.2\tablenotemark{a} & 0.54\tablenotemark{e} ; 
0.55\tablenotemark{f} ; 0.70\tablenotemark{g} & 163.4 $\times$ 
108.2\tablenotemark{d} & 36.0\tablenotemark{a} & Quadrupolar \\
Pb 6           & 56.0\tablenotemark{b} & 4.38\tablenotemark{e} ; 
4.42\tablenotemark{f} ; 4.00\tablenotemark{g} & 5.5\tablenotemark{e} & ? & 
Unclassified \\
NGC 2867 &  14.4\tablenotemark{b} & 1.84\tablenotemark{e} ; 
2.23\tablenotemark{f} ; 1.60\tablenotemark{g} & 14.4 $\times$ 
13.9\tablenotemark{d} & 18.9\tablenotemark{a} & Elliptical \\
Sand 3      &  -92.5\tablenotemark{c} & ? & -- & -- & No detectable nebulosity 
\\
\hline
\end{tabular}
$^{a}$\citet{Acker1992}; $^{b}$\citet{2013RMxAA..49...87P}; 
$^{c}$\citet{Feibelman1996}; $^{d}$\citet{Tylenda2003}; 
$^{e}$\citet{1992A&AS...94..399C}; $^{f}$\citet{Stanghellini2008}; 
$^{g}$\citet{Maciel1984}.
\end{table}

\begin{table}
\begin{center}
\caption{Spectral data sets analysed in this work.}
\label{spectra}
\begin{tabular}{llllllll}
\hline
Instrument & Data ID & Date & Resol. [$\mathrm{\AA}$] & Aperture [''] & Range 
[$\mathrm{\AA}$] & comments \\
\hline
\textbf{NGC 6905} &&&&&&\\
FUSE          	 & A1490202000  & 2000 Aug 11 & $\sim$0.06 & 30$\times$30    & 
905$-$1187 & \\
STIS + G140L 	 & O52R01020    & 1999 Jun 29 & $\sim$1.20 & 52$\times$0.5   & 
1150$-$1730 & \\
STIS + G230L	 & O52R01010    & 1999 Jun 29 & $\sim$3.15 & 52$\times$0.5   & 
1570$-$3128 & \\
\textbf{NGC 5189} &&&&&&\\
FUSE          	 & S6013001000  & 2002 Feb 28 & $\sim$0.06 & 30$\times$30    & 
905$-$1187 & \\
IUE     	 & LWR07171     & 1980 Mar 12 & $\sim$7.0  & 10$\times$20 & 
1850$-$3300 & \\ 
IUE 	         & SWP08219     & 1980 Mar 12 & $\sim$6.0  & 10$\times$20 & 
1150$-$2000 & \\ 
IUE 		 & SWP06327	& 1979 Aug 30 & $\sim$6.0  & 10$\times$20 & 
1150$-$2000 & off star\\
IUE 		 & LWR01985	& 1978 Aug 05 & $\sim$7.0  & 10$\times$20 & 
1850$-$3300 & off star\\
IUE    		 & SWP02206	& 1978 Aug 05 & $\sim$6.0  & 10$\times$20 & 
1150$-$2000 & off star\\
IUE   		 & SWP25365	& 1985 Mar 05 & $\sim$6.0  & 10$\times$20 & 
1150$-$2000 & off star\\
IUE 		 & LWP05456	& 1985 Mar 05 & $\sim$7.0  & 10$\times$20 & 
1850$-$3300 & \\
IUE 		 & LWP05457	& 1985 Mar 05 & $\sim$7.0  & 10$\times$20 & 
1850$-$3300 & off star\\
IUE 		 & SWP25363	& 1985 Mar 05 & $\sim$0.2  & 10$\times$20 & 
1150$-$2000 & bad extraction\\
IUE 		 & SWP25364	& 1985 Mar 05 & $\sim$6.0  & 10$\times$20 & 
1150$-$2000 & \\
IUE 		 & SWP41913 	& 1991 Jun 24 & $\sim$6.0  & 10$\times$20 & 
1150$-$2000 & off star\\
\textbf{Pb 6}   &&&&&&\\
FUSE                 & Z9111101000 & 2003 Feb 20 & $\sim$0.06 & 30$\times$30    
& 905$-$1187 & \\
STIS + G140L 	 & OBRZ23010   & 2012 Apr 25 & $\sim$1.20 & 52$\times$0.2   & 
1150$-$1730 & \\
STIS + G230L	 & OBRZ23020   & 2012 Apr 25 & $\sim$3.15 & 52$\times$0.2   & 
1570$-$3180 & \\
STIS + G230L	 & OBRZ23030   & 2012 Apr 25 & $\sim$3.15 & 52$\times$0.2   & 
1570$-$3180 & \\
\textbf{NGC 2867}   &&&&&&\\
FUSE                 & Z9110901000 & 2003 Feb 22 & $\sim$0.06 & 30$\times$30    
& 905$-$1187 & \\
IUE & SWP05234 & 1979 May 14 & $\sim$6.0  & 10$\times$20 & 1150$-$2000 & noisy 
\\
IUE & SWP05215 & 1979 May 12 & $\sim$6.0  & 10$\times$20 & 1150$-$2000 &  \\
IUE & LWR04510 & 1979 May 12 & $\sim$7.0  & 10$\times$20 & 1850$-$3300 &\\
IUE & LWR04518 & 1979 May 14 &$\sim$7.0  & 10$\times$20 & 1850$-$3300 &\\
IUE & SWP08984 & 1980 May 12 & $\sim$0.2  & 10$\times$20 & 1150$-$2000 & bad 
extraction\\
IUE & SWP52948 & 1994 Nov 30 & $\sim$0.2  & 10$\times$20 & 1150$-$2000 & bad 
extraction\\
IUE & LWR07736 & 1980 May 12 & $\sim$0.2  & 10$\times$20 & 1850$-$3300 & bad 
extraction\\
\textbf{Sand 3}   &&&&&&\\
STIS + G140L     & O4XH01030    & 1998 Sep 03 & $\sim$1.20 & 52$\times$2     & 
1138$-$1725 & \\
STIS + G230MB    & O4XH01020   & 1998 Sep 03 & $\sim$0.33 & 52$\times$2     & 
2716$-$2872  & \\
IUE              & SWP53882    & 1995 Feb 10 & $\sim$6.0  & 10$\times$20 & 
1150$-$1975 & \\
IUE              & LWR05765    & 1979 Oct 06 & $\sim$7.0  & 10$\times$20 & 
1860$-$3300 & \\
\hline
\end{tabular}
\end{center}
\footnotesize{SWP53882 is only used in the interval 1725$-$1851 $\mathrm{\AA}$.}
 \normalsize
\end{table}

\subsection{Description of the utilized data}
\subsubsection{FUSE data}

The FUSE instrument included four channels (LiF1, LiF2, SiC1, SiC2), each 
divided into two segments (A and B), collecting light simultaneously. Each 
channel-segment combination spans a fraction of FUSE's nominal wavelength 
coverage, with some overlap among them. All FUSE spectra used in this work were 
obtained through the $30''\times30''$ aperture, in time-tag mode. 
We used the CalFUSE v3.2.2 pipeline for data reduction, along with the FUSE IDL 
tools `CF\_edit' and `FUSE\_register'. We screened the observations for bad time 
intervals, and checked for possible drifting  of the target out of the aperture. 
The spectra of the different segments were then extracted with the 
`cf\_extract\_spectra' routine. Faulty regions of the spectra, such as those 
close to the detectors' edges and the region of the LiF 1B detector affected by 
the artefact known as `worm' (due to a shadow thrown on the detector by the 
instrument itself) were removed and the spectra of the different segments, 
co-added weighted by their errors.

In the case of CSPN Pb 6, continuum flux levels differ between FUSE and HST/STIS 
spectra, with the former being 40 per cent higher. Such divergences in flux 
levels are often related to the positioning of the target into the instrument's 
aperture or difficulties in extracting the background in low signal-to-noise 
data. A comparison between HST G140L and IUE low resolution (SWP29857) spectra 
of Pb 6 shows agreement in continuum levels within 10 per cent. Flux levels also agree between 
G140L and G230L spectra of Pb 6. According to \citet{1998stis.rept...20B}, the STIS 
$52''\times0.2''$ aperture is calibrated to a few percent accuracy, with the repeatability for point 
sources being at worse 10 per cent, with a RMS of 4.5 per cent. We thus have scaled the flux level 
of the FUSE spectrum of Pb 6 down by dividing it by 1.4 to match that of the 
STIS spectrum. 

The FUSE spectra of the central stars studied here present some strong and 
narrow emissions that do not seem to originate in the atmospheres of these 
stars, for they are much narrower than their typically broad wind lines. Some of 
these emissions are due to the planetary nebulae and some are due to airglow or 
scattered sun light, as verified by comparing the spectra obtained during the 
orbital night with the composite of all data. Except for NGC 2867, for which the 
signal to noise ratio of the available FUSE spectra allowed us to use night only 
events, thus minimizing airglow contamination, both orbital night and day 
time-tag events were used in the stellar spectra of the sample objects.

The central stars of NGC 2867 and NGC 5189 have 3 FUSE observations each. We have used those with the 
highest signal-to-noise ratio. While the plot previews available 
at the MAST website show some variation of flux levels and line profiles between different data sets, the careful treatment of the data, as 
described above, aiming at mitigating the effects of channel drift, airglow, faulty centroid positioning, worms 
and dead zones, strongly diminishes the disagreements between the spectra obtained in different observations, 
which are now on the same level as those seen comparing spectra from the same observation, but from different channels, suggesting that instrumental effects plus others such as airglow (which can be minimized by the use of night only time-tag events, but is never completely absent) may be responsible 
for the remaining small variations in line profiles and flux levels.

\subsubsection{IUE data}

\begin{figure}
\centering
\includegraphics[scale=0.435]{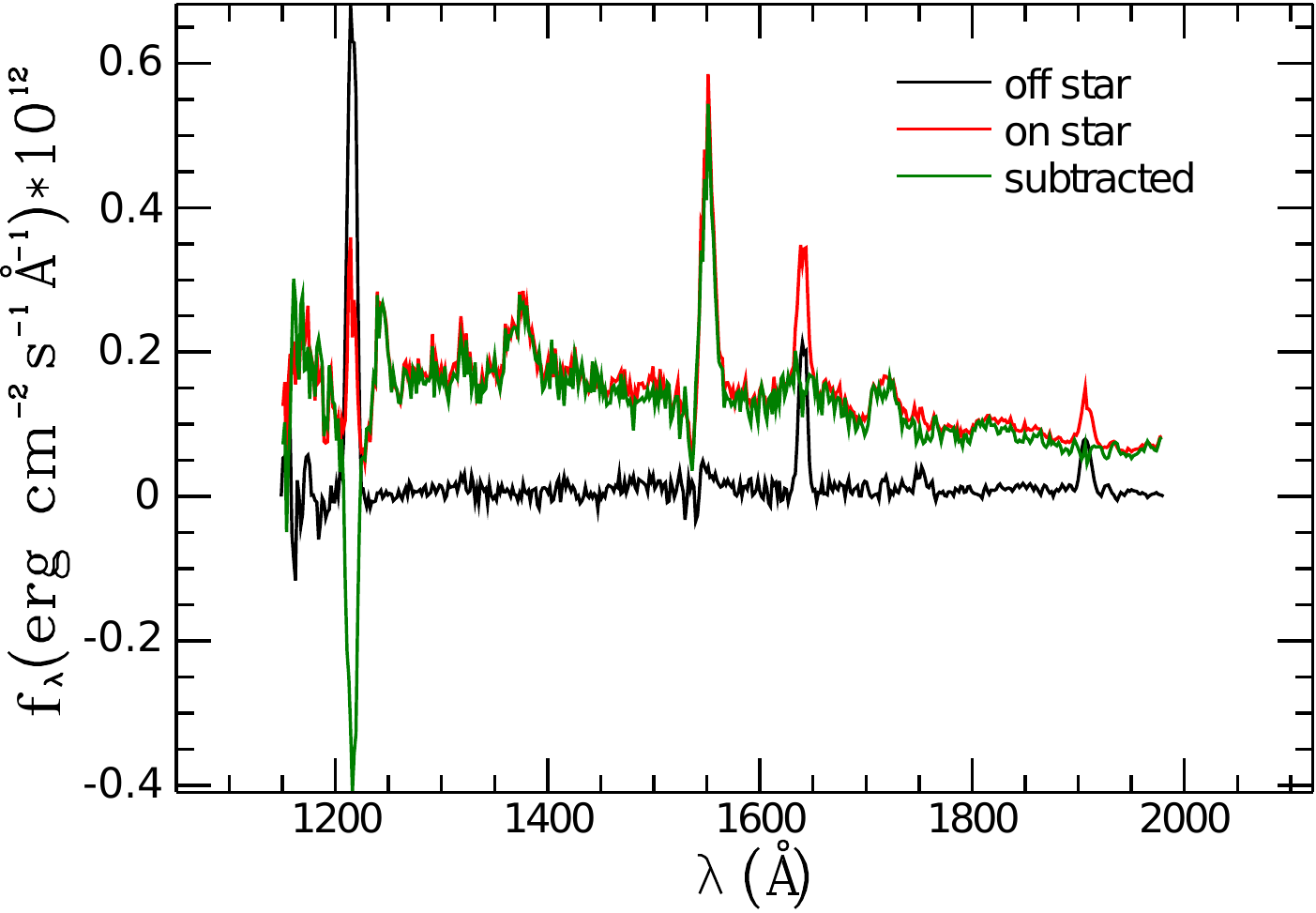}
\caption{SWP25364 (red) and SWP25365 (black) low-resolution IUE spectra. The 
result of subtracting SWP25365 from 
SWP25364 is shown in green.}\label{subtraction}
\end{figure}

Despite the fact that MAST lists all IUE spectra for NGC 5189 under the category 
70, which is described there as observation of nebula plus central star, our 
analysis of the spectra indicate that many of the exposures actually missed the 
central star, as in the case of SWP25363 - the only high resolution IUE spectrum 
available for NGC 5189. In the 2D high resolution SWP25363 spectral image, 
spatially extended nebular emission lines and only a very faint continuum are 
visible. The continuum is not centred, but at the edge of the illuminated 
region, which is evidence that flux was partially missed. The pipeline seems to 
have misplaced the extraction slit and part of the resulting net flux is below 
zero. Nonetheless, the spectral resolution of 0.2 $\mathrm{\AA}$ allows us to 
estimate the width of the He\,{\sc ii} $\lambda$ 1640.4 $\mathrm{\AA}$ line as $\approx$0.6 
$\mathrm{\AA}$, consistent with a nebular origin. We 
thus cannot establish from the echelle spectrum whether there is also stellar He\,{\sc ii} $\lambda$ 1640.4 $\mathrm{\AA}$ emission. Fig. \ref{subtraction} shows IUE 
SWP25364 and SWP25365 low-resolution spectra, which were taken at the same 
pointing coordinates and with the large aperture (10'' $\times$ 20''), 
but at different position angles. The latter seems to have at least partially 
missed the central star. The difference between the two spectra exposes the 
nebular origin of the C\,{\sc iii]} $\lambda$ 1909 $\mathrm{\AA}$ line and indicates 
that at least part of the He\,{\sc ii} $\lambda$ 1640.4 $\mathrm{\AA}$ emission seen in 
SWP25364 is of nebular origin. Therefore, we will regard NGC 5189's He\,{\sc ii} 
$\lambda$ 1640.4 $\mathrm{\AA}$ emission as an upper limit to the stellar line 
and leverage on other spectral lines to constrain the stellar parameters. 
Since there are no high-resolution spectra of NGC 5189's central star in the 
region between 1200 and 3200 $\mathrm{\AA}$, we rely on the low resolution 
data. For the present analysis of NGC 5189, we use SWP08219 and 
LWR07171, for being better centred and less noisy than the SWP25364 and 
LWP05456 spectra, respectively.

The available IUE low resolution spectra of NGC 2867 seem to be dominated by 
nebular emissions, such that P-Cygni profiles cannot be distinguished. The 
inspection of the spectral images of SWP05215 and LWR04510 shows well defined, 
well centred and narrow continua. Since only the strongest emission lines in the 
SWP05215 spectrum appear saturated and because its continuum has a higher signal than that of the noisy SWP05234 spectrum, we chose to show the former instead of 
the latter in this work, for it allows a more precise determination of the slope 
of the continuum, which will be used in the normalization of the observed 
spectra and to ascertain the colour excess. We will not use it for line strength 
analysis, except as an upper limit for stellar emission in some weaker lines. 
The existing high resolution IUE spectra of this star exhibit again very faint 
continua (in the two-dimensional spectra) and negative flux values (in the extracted spectra), indicating 
again extraction problems. No obvious stellar lines are apparent. Narrow and 
strong C\,{\sc iv} $\lambda \lambda$ $1548.2$, $1550.8$ $\mathrm{\AA}$, He\,{\sc ii} $\lambda$ 
1640.4 $\mathrm{\AA}$ and C\,{\sc iii]} $\lambda$ 1909 $\mathrm{\AA}$ nebular emission 
lines are, however, readily discernible in the high resolution spectra.


\subsubsection{HST data}

HST/STIS gratings are designed for spatially resolved spectroscopy using a long 
slit, exploiting HST's high spatial resolution capabilities. The HST/STIS 
spectra of CSPNe NGC 6905, Pb 6, and Sand 3 thus minimize nebular contamination 
of the stellar extracted spectra - in comparison with IUE spectra, for example. 
The inspection of the 2D G140L and G230L spectra of NGC 6905 and Pb 6 reveals 
extended nebular emission around $\approx$1909 (very faint in NGC 6905), 
1550 and 1640 $\mathrm{\AA}$, as seen on the logarithmic scale plots in the 
supplementary figures, online. Other faint extended emissions are seen in 
the 2D G140L spectrum of Pb 6 around $\approx$1240, 1400 and 1485 $\mathrm{\AA}$. 
The 2D Sand 3 G140L spectrum does not exhibit evidence of nebular emissions, in 
line with the idea of its nebula having already dissipated \citep{Barlow1982}.

In neither 2D STIS spectra there is evidence of extended nebular emission close to the structure 
associated by \citet{Feibelman1996} with nebular [Mg\,{\sc v}] $\lambda \lambda$ 
$1324.5$ $\mathrm{\AA}$ in 
the low resolution IUE spectra of Sand 3 and marked in Fig. \ref{UVspectra}, even though the extracted stellar spectra 
of all three objects observed with STIS show the feature, 
which, together with the evidence of Sand 3's nebula having dispersed, points towards a stellar origin for this spectral feature. 
Similarly, \citet{Feibelman1996a} suggested the nebular He\,{\sc ii} $\lambda$ 2253 $\mathrm{\AA}$ and numerous [Ne\,{\sc v}] lines between 
2235 and 2265 $\mathrm{\AA}$ seen in the high resolution IUE spectra of NGC 6905 as 
the sources of the wide emission feature seen in lower resolution spectra of 
this region. The origin of this spectral feature, however, seems to be at least partially stellar, since 
the strong feature appears not only in the low resolution IUE spectra of Sand 3 
- despite its dispersed nebula not producing a detectable C\,{\sc iii]} $\lambda$ 1909 
$\mathrm{\AA}$ emission - but also in the 1D STIS spectra of NGC 6905 and Pb 6, 
whose 2D G230L spectra do not show evidence of extended nebular emission in this region. 

The agreement between the medium resolution 
($\approx$0.33 $\mathrm{\AA}$) HST/STIS G230MB observed spectrum of Sand 3's O\,{\sc v} 
$\lambda \lambda$ $2781.0$, $2787.0$ $\mathrm{\AA}$ line and the low resolution IUE 
spectrum of the same star indicates that this strong and wide line is from stellar origin. This was also verified from the inspection of the 2D G230MB spectrum 
of Sand 3 and of the 2D G230L spectra of NGC 6905 and Pb 6, which do not show 
extended emissions in this region.

\section{Modelling}
\label{sec:thegrid}

\subsection{Stellar models}

The spectra of these very hot, H-poor CSPNe are dominated by broad, strong 
emission lines, which are signatures of their dense, high velocity stellar 
winds. Spectra of hot objects with detectable winds require modelling with 
non-LTE stellar atmosphere codes that account for expanding atmospheres. In this 
work, we used models computed with CMFGEN 
\citep{1998ApJ...496..407H,2005ApJ...627..424H}, which is a non-LTE, 
non-hydrodynamic code that considers a stationary mass loss in a spherically 
symmetric geometry, line blanketing (through the super-level approximation) and 
wind clumping (through the filling factor approach). The detailed workings of 
the code are described in the references above. The fundamental stellar 
parameters include stellar temperature ($T_{\ast}$) and radius ($R_{\ast}$), 
defined at an optical depth of $\tau=$20, mass-loss rate ($\dot{M}$), terminal 
velocity of the wind ($v_{\infty}$), velocity law, and elemental abundances. We 
assume a standard velocity law, $v(r)=v_{\infty}(1-r_{0}/r)^{\beta}$, where 
$r_{0} \approx R_{\ast}$ and $\beta = 1$, and an exponential law for the clumping filling factor 
(which is the fraction of the wind volume occupied by clumps), given by 
$\textit{f}(v) = \textit{f}_{\infty} + (1 - 
\textit{f}_{\infty})\exp{(-v(r)/v_{clump})}$, where $v_{clump}=200$ km s$^{-1}$ 
is the velocity beyond which the wind is no longer smooth and the clumping increases until the terminal value of $\textit{f}_{\infty}=0.1$ is 
reached.

The analysis leverages on the extensive grid of synthetic spectra computed by 
\citet{kelleretal2011PORT}, covering parameter values typical of H-poor CSPNe 
and approximately following the evolutionary tracks from 
\citet{2006A&A...454..845M}.
The grid allows for a uniform and systematic comparison of the spectral 
features and facilitates line identification and the determination of the main 
stellar parameters. The grid models adopt a constant abundance pattern: 
X$_{\mathrm{C}}=0.45$, X$_{\mathrm{He}}=0.43$, X$_{\mathrm{O}}=0.08$, 
X$_{\mathrm{N}}=0.01$, X$_{\mathrm{Ne}}=0.02$ in mass fractions, an iron 
abundance a factor of $100$ lower than the solar value, and solar abundances for 
all other elements included into the calculations. Within the parameter space of 
[WCE] CSPNe, the grid models have temperatures of $\approx$ 100,000, 125,000, 
150,000, 165,000, and 200,000 K. The detailed emerging flux was calculated 
adopting a varying microturbulence velocity described by 
$\xi(r)=\xi_{min}+(\xi_{max}-\xi_{min})v(r)/v_{\infty}$, with $\xi_{min}=10$ km 
s$^{-1}$ near the photosphere and $\xi_{max}=50$ km s$^{-1}$. The grid models 
were computed including many ionic species not considered in previous analyses 
and are described in \citet{kelleretal2011PORT}, 
\citet{Keller2012a,KellerESOinpress}.\footnote{The grid covering the parameter 
regime of the [WC] CSPNe is available on-line at 
\url{http://dolomiti.pha.jhu.edu/planetarynebulae.html}, where the user will 
find tables listing the available models and individual pages for every model, 
with information about the stellar parameters adopted, the ions considered and 
the synthetic spectra, as well as figures comparing the different models and 
identifying the spectral lines predicted.} Additional ad-hoc models were 
computed for each object with varying elemental abundances and microturbulence 
velocities, and including additional less abundant ions. The ions included 
in the final best-fitting models, together with the number of levels and 
superlevels adopted in the calculations, are given in Table \ref{tab:ions}.

\begin{table}
\begin{center}
 \scriptsize
\caption{Ion superlevels and levels of the best-fitting final models for the 
central stars of NGC 6905, Sand 3, Pb 6, NGC 2867, and NGC 5189.}
\begin{tabular}{cccccccccccc}
\hline
Element &         I   &         II  &    III  &   IV         &   V           &   
VI          &   VII            &   VIII             &   IX            &   X      
     &   XI \\
\hline
He      & 40,45       & 22,30       &   1     &              &               &   
            &                  &                    &                 &          
     &  \\
C       &             &             &         &  49,64       &   1           &   
            &                  &                    &                 &          
     &  \\
N       &             &             &         &              & 13,21         &   
1           &                  &                    &                 &          
     &  \\  
O       &             &             &         &29,48\tablenotemark{a}& 58,163    
    & 41,47         &    1             &                    &                 &  
             &  \\  
Ne      &             &             &         &45,355\tablenotemark{a}& 37,166   
    & 36,202        &  38,182          &  24,47             &   1             &  
             &  \\ 
Na      &             &             &         &              &               & 
52,452        &  37,251          &  72,214            & 27,71           &  1     
       &  \\ 
Mg      &             &             &         &              & 43,311        & 
46,444        &  54,476          &    1               &                 &        
       &  \\ 
Si      &             &             &         &22,33 & 33,71\tablenotemark{a}& 
33,98\tablenotemark{a}&    1\tablenotemark{a}&                    &              
   &               &  \\ 
P       &             &             &         &              & 16,62         &  
1            &                  &                    &                 &         
      &  \\ 
S       &             &             &         &              &               & 
28,58         &    1             &                    &                 &        
       &  \\  
Ar      &             &             &         &              &               & 
30,205        &  33,174          &  57,72             &    1            &        
       &  \\ 
Ca      &             &             &         &              &               & 
47,108\tablenotemark{a}&  55,514\tablenotemark{a}&  54,445\tablenotemark{a}& 
35,367\tablenotemark{a}& 31,79\tablenotemark{a}& 1\tablenotemark{a} \\
Fe      &             &             &         &              &               & 
44,433\tablenotemark{a}&  41,252         &  53,324            & 52,490          
& 43,210        & 1    \\
Co      &             &             &         &              &               &   
            &  45,1000         &  50,1217           & 24,355          &   1      
     &  \\ 
Ni      &             &             &         &              &               &   
            &  37,308          &  113,1000          & 75,1217         &   1      
     &  \\

\hline
\end{tabular}
\label{tab:ions}
\tablenotetext{a}{These ionic species are not present in the best-fitting final 
models for the central stars of NGC 5189, Pb 6, and NGC 2867.}
 \normalsize
\end{center}
\end{table}

\subsection{Effects of the interstellar medium}

\begin{figure}
\centering
\includegraphics[scale=0.5]{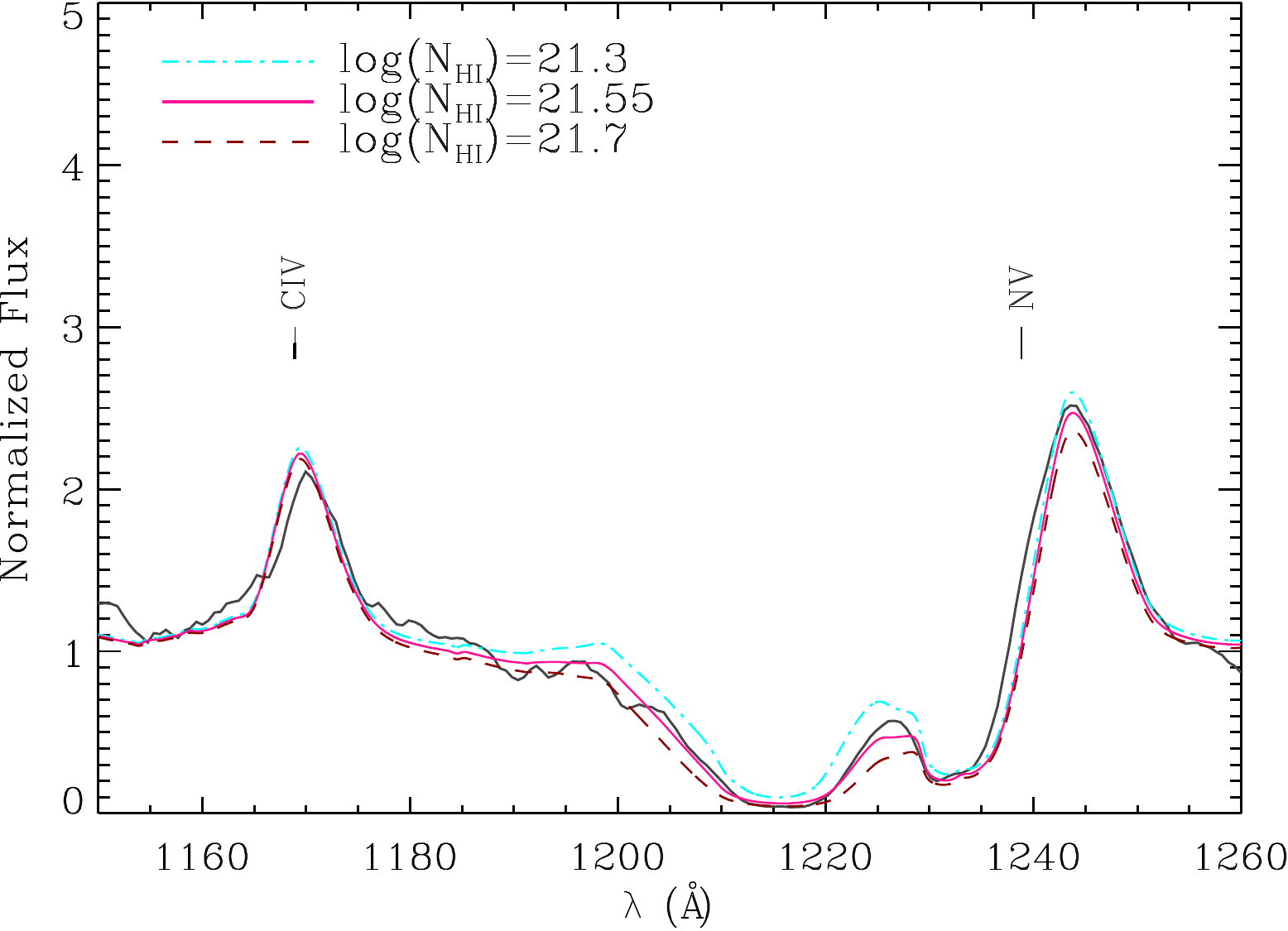}
\caption{The interstellar Ly\,$\alpha$ absorption seen in the observed spectra 
of Sand 3 
(continuous black line) is compared to our model, to which we applied the effect 
of different values of neutral hydrogen column density. 
The model shown has $T_{\ast}=150$ kK, $R_{t}=$9.4 $R_{\odot}$ and 
$X_{N}=0.07$}\label{sand3Lyalpha}
\end{figure}

In order to separate stellar features from interstellar absorptions due to 
neutral and molecular hydrogen along the line of sight, which greatly affect the 
far-UV spectra of these objects, we modelled them as described in 
\citet{Herald2002,2004ApJ...609..378H,Herald2004a}, assuming a temperature of the interstellar gas of 100 K, which is a typical value for the ISM, and determined the column densities. The derived interstellar parameters 
are given in Table \ref{tab:is}, where $N(H$\,{\sc i}$)$ and $N(H_{2})$ are the neutral 
and molecular hydrogen column densities, respectively. Fig. \ref{sand3Lyalpha} 
shows the interstellar Ly\,$\alpha$ absorption seen in the observed spectra of 
Sand 3 and illustrates the fact that accounting for atomic hydrogen ISM 
absorption can affect the modelling of the N\,{\sc v} $\lambda \lambda$ 1238.8, 1242.8 
$\mathrm{\AA}$ line. We did not attempt to model the ISM Ly\,$\alpha$ 
absorption in low resolution IUE spectra of CSPNe NGC 2867 and NGC 5189.

\begin{table}
\begin{center}
 \scriptsize
\caption{Interstellar parameters adopted in the modelling.}
\begin{tabular}{lccccc}
\hline
Star & $\log N(H$\,{\sc i}$)$\tablenotemark{b} & $\log N(H_{2})$\tablenotemark{c} & 
$T_{gas}$ [K]\tablenotemark{d} & $E_{B-V}$ [mag] & R$_{v}$\\
\hline
Pb 6 & 21.50 & 20.20 & 100 & 0.25$\pm^{0.05}_{0.05}$ & 3.1\\
NGC 5189 & $-$ & 20.30 & 100 & 0.23$\pm^{0.07}_{0.03}$ & 3.1\\
NGC 2867 & $-$ & 19.90 & 100 & 0.18\tablenotemark{e} & 3.1\\
NGC 6905\tablenotemark{a} & 21.00 & 19.50 & 100 & 0.17$\pm^{0.13}_{0.07}$ & 3.1 
\\
Sand 3 & 21.55 & $-$ & 100 & 0.40$\pm^{0.05}_{0.05}$ & 3.1\\
\hline
\end{tabular}
\label{tab:is}
\tablenotetext{a}{From \citet{kelleretal2011PORT}.}
\tablenotetext{b}{From Ly\,$\alpha$. $N(H$\,{\sc i}$)$ is given in units of 
cm$^{-2}$.}
\tablenotetext{c}{From the absorption lines in the FUSE range. $N(H_{2})$ is 
given in units of cm$^{-2}$.}
\tablenotetext{d}{Assumed.}
\tablenotetext{e}{Derived from the extinction quantity taken from 
\citet{2002ApJS..138..279M}.}
 \normalsize
\end{center}
\end{table}

\section{Spectral analyses}
\label{sec:SpecAnalyses}

As the first step in the spectral analysis of the stellar lines, we estimated 
the terminal velocity of the wind from the short-wavelength edge of the strong 
P-Cygni profile of C\,{\sc iv} $\lambda$ 1548.2 $\mathrm{\AA}$ and also from the long 
wavelength component of the O\,{\sc vi} $\lambda \lambda$ $1031.9$, $1037.6$ 
$\mathrm{\AA}$ doublet whenever both or any of these lines were available, 
assuming a turbulence of $10$ per cent the terminal velocity of the wind.

The spectra of [WCE] CSPNe are particularly poor in spectral lines of different 
ionization stages of the same element, and for this reason, the absence or 
presence of structures due to ions of different ionization potentials and the 
overall appearance of the spectra are also used to constrain temperature. As was 
shown by \citet{kelleretal2011PORT}, oxygen is one of the few elements producing 
strong lines of different ionization stages in the spectra of [WCE] CSPNe, with 
several O\,{\sc vi} and two O\,{\sc v} lines visible in the synthetic spectra throughout the 
temperature range of [WCE] CSPNe. As found by \citet{kelleretal2011PORT}, one 
important aspect about these O\,{\sc v} lines, which can influence the determination of 
the stellar temperature, is their considerable increase in strength, in the 
synthetic spectra, when less abundant ions are included into the calculation, 
especially Mg, Na, Co, and Ni. These ions are not included in the grid models, 
because of computational limitations, but we include them in our final 
best-fitting models, as we show below. 

The most important line diagnostics we rely upon when constraining the surface 
temperatures of these stars are the Ne\,{\sc vii} $\lambda$ 973.3 $\mathrm{\AA}$ and O\,{\sc vi} $\lambda \lambda$ 1031.9, 1037.6 $\mathrm{\AA}$ P-Cygni profiles and the O\,{\sc v} 
$\lambda$ $1371.3$ $\mathrm{\AA}$ and $\lambda \lambda$ $2781.0$, $2787.0$ 
$\mathrm{\AA}$ lines. The Ne\,{\sc vii} $\lambda$ 973.3 $\mathrm{\AA}$ line is a useful 
temperature diagnostic, because its synthetic profile is substantially weakened 
by the decrease in temperature in the [WCE] parameter regime and because it 
shows little sensitivity to mass loss at model temperatures of 100,000 and 
125,000 K, in the parameter regime covered by our model grid (Fig. 
\ref{FUSEspectra}). At these temperatures, it also shows little sensitivity to 
neon abundance, as shown by \citet{kelleretal2011PORT}. A similar behaviour with 
temperature is observed in the synthetic O\,{\sc vi} $\lambda \lambda$ 1031.9, 1037.6 
$\mathrm{\AA}$ doublet, whose strength also increases towards models with 
T$_{\ast}=$165,000 K and which is not particularly sensitive to variations in 
mass loss at model temperatures of 100,000 and 125,000 K. Thus, the strong Ne\,{\sc vii} $\lambda$ 973.3 $\mathrm{\AA}$ and O\,{\sc vi} $\lambda \lambda$ 1031.9, 1037.6 
$\mathrm{\AA}$ P-Cygni profiles observed in all the objects with FUSE spectra 
shown here, immediately point towards models of T$_{\ast} = $150,000 K or 
165,000 K, with the latter allowing for even stronger line profiles, 
concomitantly with good fits to the remaining lines. 

One of the most striking differences among the observed spectra of the studied 
stars, which are otherwise very similar, is the strength of the O\,{\sc v} lines. The 
central star of NGC 6905 and Sand 3 show much stronger lines than is seen in the 
spectra of NGC 5189 and of Pb 6, or even than is present on the nebula-dominated 
IUE spectra of NGC 2867. This characteristic, along with the line 
appearance, can further constrain temperature: the appearance of the O\,{\sc v} 
$\lambda$ $1371.3$ $\mathrm{\AA}$ line profile also excludes models with 
T$_{\ast} =$100,000 and 125,000 K, since none of the observed line profiles show 
the deep P-Cygni absorption components that are seen in these models. Also, 
although the grid models with T$_{\ast} = $150,000 K and 165,000 K show weak O\,{\sc v} 
lines for mass-loss rates that allow reasonable fits of the overall observed 
spectra, their synthetic profiles do become much stronger through the addition 
of further ions into the calculations, with the T$_{\ast} = $150,000 K models 
producing O\,{\sc v} profiles considerably stronger than models with T$_{\ast} = 
$165,000 K. We thus concluded that the central star of NGC 6905 and Sand 3 are 
cooler than the other three objects and best fit by models of T$_{\ast} = 
$150,000 K, while the central stars of NGC 2867, NGC 5189, and Pb 6 best-fitting 
models have T$_{\ast} = $165,000 K.

Hotter models, with T$_{\ast} = $200,000 K, were also computed by us. However, 
they do not fit well the observed spectra: at the same time, the models show too 
weak O\,{\sc vi} $\lambda \lambda$ 1031.9, 1037.6 $\mathrm{\AA}$ lines and too strong 
He\,{\sc ii} $\lambda$ 1640.4 $\mathrm{\AA}$ and C\,{\sc iv} $\lambda\lambda$ 1107.6, 1107.9 
$\mathrm{\AA}$, $\lambda\lambda$ 1168.8, 1169.0 $\mathrm{\AA}$ and 
$\lambda \lambda$ $1548.2$, $1550.8$ $\mathrm{\AA}$ lines, such that the 
discrepancy is too big to be solved by different abundance patterns. The 200,000 
K models also predict Ne\,{\sc vii} $\lambda$ 973.3 $\mathrm{\AA}$ lines with shallower 
and narrower absorption components than what is seen in the observations and no 
O\,{\sc v} emission lines.

Another factor that influences the appearance of the spectral lines in these 
stars is how dense their wind is. \citet{1989A&A...210..236S} have introduced 
the so called transformed radius, defined as 
$R_{\mathrm{t}}=R_{\ast}[(v_{\infty}/2500\ \mathrm{km}\ \mathrm{s}^{-1}) / 
(\dot{M}/10^{-4}\ \mathrm{M}_{\odot}\mathrm{yr}^{-1})]^{2/3}$. Models of the 
same transformed radius that also have identical temperature, composition, and 
terminal velocity - that is, models where $R_{\ast} / \dot{M}^{2/3}$ is the same 
- are known to have similar ionization structure and temperature stratification 
and to lead to very similar wind line spectra \citep{1993A&A...274..397H}. In 
this work, we will constrain the transformed radius rather than the mass-loss 
rate, because deriving mass-loss from $R_{t}$ depends on a measurement of the stellar radius, which depends on the distance and distances to Galactic CSPNe 
are commonly not well known. On top of this, the absence of strong absorption 
lines not contaminated by wind emissions in the spectra of [WCE] stars prevents 
us from measuring $\log g$.

The overall strength of all available wind lines was used by us to constrain the 
transformed radius. Among them, the He\,{\sc ii} $\lambda$ 1640.4 $\mathrm{\AA}$ - 
whenever not strongly affected by nebular emission - is especially useful due to 
its little sensitivity to temperature in the parameter regime covered by the 
model grid for [WCE] stars. The behaviour of the synthetic spectral lines 
apparent in the spectra of [WCE] stars with stellar parameters was extensively 
described in \citet{kelleretal2011PORT}. 

After having used the grid of synthetic spectra to constrain stellar temperature 
and transformed radius, we then extended the analysis by exploring various 
abundances for the main elements and different microturbulence velocities, as 
well as by including additional elements into the calculations. The temperature steps of our model grid, in the [WCE] regime, 
are small enough that subsequent models do not show 
strong variations of line diagnostics. That is the reason for the non-homogeneous temperature 
intervals in our model grid. In our ad-hoc analysis, temperature was therefore not 
varied further. At each likely temperature, a number of models was calculated with different elemental abundances, as to ensure that 
different abundance patterns would not affect the constraining of temperature.

\begin{figure}
\centering
\includegraphics[scale=0.435]{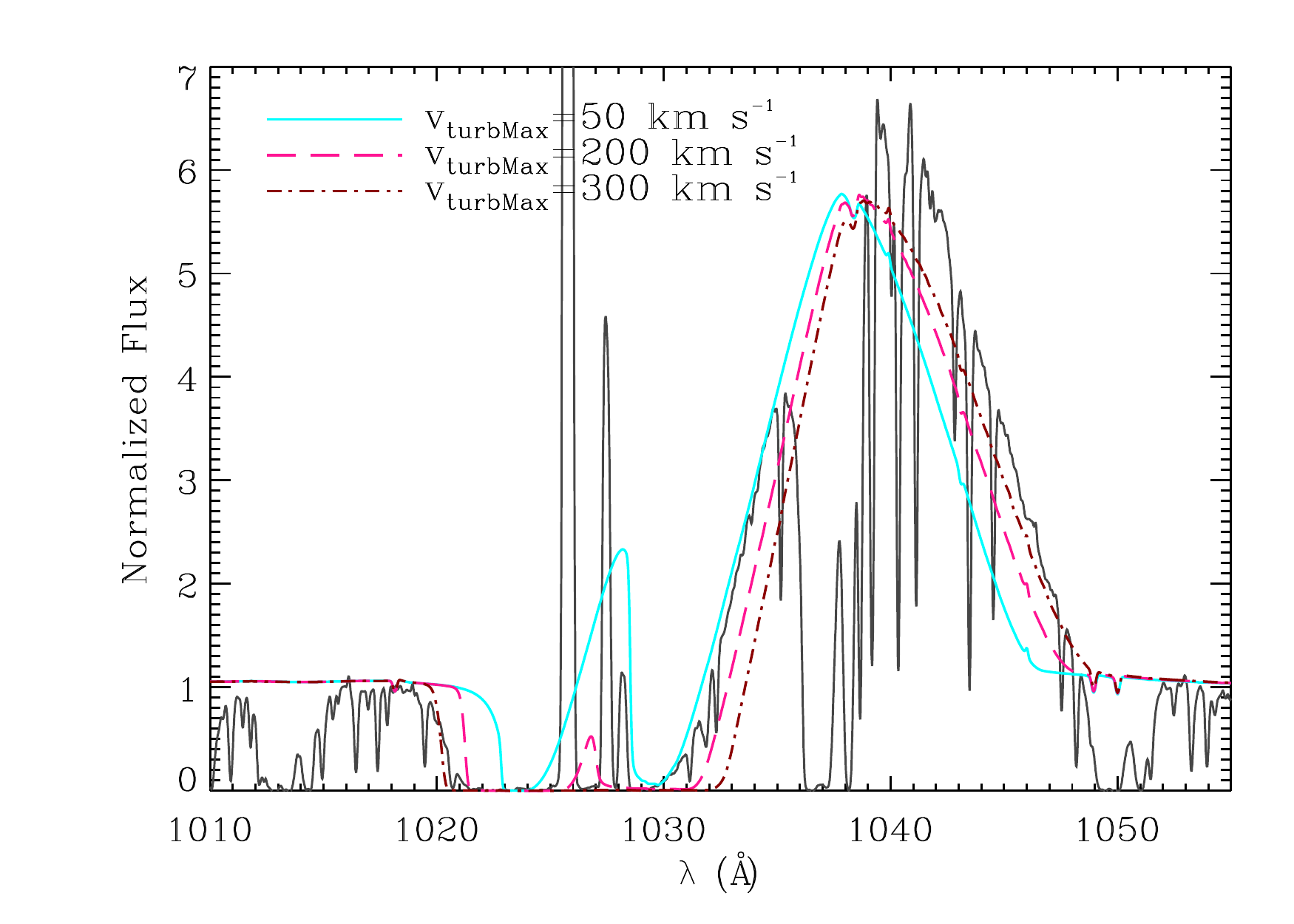}
\caption{CSPN NGC 5189 observed O\,{\sc vi} $\lambda \lambda$ $1031.9$, $1037.6$ 
$\mathrm{\AA}$ line (black continuous line) is compared with a grid model with 
$T_{\ast}=165$ kK, $R_{t}=9.9R_{\odot}$, 
$v_{\infty}=2500$ 
km s$^{-1}$ recomputed to account for different microturbulence 
velocities.}\label{vturbNGC5189}
\vspace{0.5cm}
\end{figure}

For all objects 
studied here, we found that a microturbulence velocity higher than that adopted 
in the grid models (Table \ref{tab:bestfitmodels}) improves the fits to the O\,{\sc vi} $\lambda \lambda$ 1031.9, 
1037.6 $\mathrm{\AA}$ P-Cygni profile, as shown in Fig. \ref{vturbNGC5189}, and 
also to the C\,{\sc iv} $\lambda \lambda$ $1548.2$, $1550.8$ $\mathrm{\AA}$ line.

\subsection{Results for individual objects}

The observed spectra and final best-fitting models for the individual objects 
are shown in Figs. \ref{allfits_FUV}, \ref{allfits_UV}, and \ref{allfits_NUV}. 
The parameters of our best-fitting models are given in Table 
\ref{tab:bestfitmodels}. The uncertainties in stellar temperature are given in 
the sections on the individual objects. Whenever far-UV, UV, and near-UV spectra 
are available for an object, we estimate the uncertainty in $T_{\ast}$ to be 
roughly half the interval between the values adopted in our grid of synthetic 
spectra. When, as is the case of NGC 2867 and Sand 3, not all the necessary line 
diagnostics are available, the uncertainties in temperature are deemed larger. 
The uncertainties in the mass-loss rate are dominated by the uncertainties in 
the distance and in the clumping filling factor, which was assumed a fixed 
parameter throughout this work. We can typically constrain $R_{t}$ within $\pm$ 
2.0 R$_{\odot}$. Assuming the stellar parameters of the best-fitting models of 
the various sample objects, we estimate the uncertainties of the carbon and 
helium mass fraction are about $\pm0.10$, as illustrated in Fig. 
\ref{carbonuncertainty}, and the uncertainties of the oxygen and nitrogen mass 
fractions to be about $\pm0.06$ (Fig. 17 from \citet{kelleretal2011PORT} 
illustrates the impact of different oxygen abundances on the synthetic spectra) 
and $\pm0.01$ (Fig. \ref{nitrogenuncertainty}), respectively. We have used the 
far-UV Ne\,{\sc vii} and Ne\,{\sc vi} features in the FUSE spectra to determine the 
neon mass fractions. However, as we discuss further below, Ne features appear in 
the synthetic spectra at longer wavelengths, which are not observed, perhaps due to uncertainties in the atomic data. The Ne\,{\sc vii} $\lambda$ $973.3$ 
$\mathrm{\AA}$ line, however, has been used to infer enhanced neon abundances 
in PG 1159 stars by \citet{2005ApJ...627..424H} and \citet{2004A&A...427..685W}. 
The atomic data used for the $\lambda$ $973.3$ $\mathrm{\AA}$ Ne\,{\sc vii} line are 
described in \citet{2005ApJ...627..424H}.  

The observed spectra shown in Figs. \ref{allfits_UV} and \ref{allfits_NUV} were 
obtained with different instruments and thus, differ in spectral resolution. 
Sand 3, NGC 6905, and Pb 6 STIS spectra were obtained with different apertures 
and thus display different line spread functions (LSFs) as well. The synthetic 
spectra were convolved with the corresponding LSFs when being compared to STIS 
observations and with Gaussians of appropriate FWHM when compared to IUE and 
FUSE data.

\subsubsection{CSPN NGC 6905 and Sand 3}

CSPN NGC 6905 and Sand 3 are the two stars for which we inferred lower 
temperatures: $T_{\ast}=150\pm^{8}_{13}$ kK for NGC 6905 and 
$T_{\ast}=150\pm^{8}_{25}$ kK for Sand 3, the difference in error bars being due 
to the lack of FUSE spectra available for Sand 3, which deprives the analysis of 
the far-UV O\,{\sc vi} and Ne\,{\sc vii} P-Cygni profiles, which are the line diagnostics that 
most clearly separate models of 125,000 and 150,000 K. We determined terminal 
velocities of the wind of $v_{\infty}=$2000$\pm 200$ and 2200$\pm 200$ km 
s$^{-1}$ respectively. Their STIS spectra are very similar and both stars show O\,{\sc v} lines much stronger than in all other sample objects. Two of 
the most conspicuous differences between the spectra of these two stars are the 
strong N\,{\sc v} $\lambda \lambda$ 1238.8, 1242.8 $\mathrm{\AA}$ P-Cygni profile seen 
in Sand 3's spectrum and absent from NGC 6905's and the even stronger O\,{\sc v} 
$\lambda \lambda$ $2781.0$, $2787.0$ $\mathrm{\AA}$ line observed in Sand 3. 
From the absence of a N\,{\sc v} $\lambda \lambda$ 1238.8, 1242.8 $\mathrm{\AA}$ line 
profile in the observed spectra of NGC 6905, we derived, in 
\citet{kelleretal2011PORT}, an upper limit to its N mass fraction of 
$5.5\times10^{-4}$, while for Sand 3, a much higher abundance of $X_{N}=0.07$ 
was found. For both stars we found the same oxygen mass fraction, 
$X_{O}=0.08\pm0.06$. Although neither best-fitting model for either star 
reproduces the observed intensity of the O\,{\sc v} $\lambda \lambda$ $2781.0$, 
$2787.0$ $\mathrm{\AA}$ line, our best-fitting model for Sand 3 predicts a 
stronger line than our model for CSPN NGC 6905, which is due to the lower value 
of transformed radius adopted in the former model ($R_{t}=9.4\pm2.0$ R$_{\odot}$ 
for Sand 3 and $R_{t}=10.7\pm2.0$ R$_{\odot}$ for NGC 6905). Because no FUSE 
spectra are available for Sand 3, we simply kept the Ne abundance used in the 
grid models. 

Carbon and helium mass fractions derived by us for CSPN NGC 6905 ($X_{C}=0.45 
\pm 0.10$, $X_{He}=0.45 \pm 0.10$) and Sand 3 ($X_{C}=0.55 \pm 0.10$, $X_{He}=0.28 
\pm 0.10$) are quite different, which illustrates the fact that the overall 
appearance of [WCE] spectra is mainly governed by temperature and mass loss, and 
He and C abundances play only a secondary role, with big changes in 
abundance producing minor changes in line strength.   

Optical and low resolution IUE spectra of Sand 3 were previously analysed by 
\cite{1997A&A...320...91K}, 
using non-LTE, homogeneous, spherically symmetric expanding atmosphere models, 
which included helium, carbon, 
oxygen and nitrogen in the calculations. They found $v_{\infty}=2200$ km/s, 
$T_{\ast}=140000$ K, and $R_{t}=3.0$ R$_{\odot}$. As for the elemental mass 
fractions, they derived $X_{He}=0.615$, $X_{C}=0.26$, $X_{O}=0.12$ and 
$X_{N}=0.005$. Values of $v_{\infty}$, $T_{\ast}$, and $X_{O}$ are in agreement 
with those derived in this work within the error bars. $X_{He}$, $X_{C}$, and 
$X_{N}$ are in strong disagreement. In a previous work, 
\citet{Barlow1982} derived He, C, and O mass fractions that are close to the 
values derived by us. A previous analysis of CSPNe NGC 6905 was discussed in 
\citet{kelleretal2011PORT}.

\subsubsection{CSPN NGC 2867}

The O\,{\sc vi} $\lambda \lambda$ 1031.9, 1037.6 $\mathrm{\AA}$ P-Cygni profile of CSPN 
NGC 2867 is clearly narrower than those observed in NGC 5189 and Pb 6 spectra 
and thus we derived $v_{\infty}=$2000$\pm 250$ km s$^{-1}$ for the central star 
of NGC 2867. If indeed the nebula-dominated low resolution IUE spectra of NGC 
2867 contains the stellar flux, we can set an upper limit to the strength of the O\,{\sc v} 
lines and thus better constrain the temperature. Since no strong O\,{\sc v} lines are 
seen, models with stellar temperatures of 165,000 K would be favoured. If not, 
we can only leverage on FUSE spectra to discern between model temperatures. 
However, this spectral region does not show remarkable differences between 
models of 150,000 and 165,000 K, although it rules out models of $T_{\ast} \le$ 
125,000 K which do not produce the strong Ne\,{\sc vii} and O\,{\sc vi} P-Cygni profiles 
observed. Models of 165,000 K do produce, however, somewhat stronger Ne\,{\sc vii} and 
O\,{\sc vi} lines, while concomitantly producing reasonable fits for the other 
available lines, than models of 150,000 K, in which these lines appear somewhat 
weaker than observed. We thus derived $T_{\ast}=165\pm^{18}_{20}$ kK for the 
central star of NGC 2867, $R_{t}=8.5\pm2.0$ R$_{\odot}$, $X_{C}=0.25 \pm0.10$, 
$X_{He}=0.60 \pm 0.10$ (even though we did not attempt a fit of the He\,{\sc ii} 1640.4 
$\mathrm{\AA}$ line due to the heavy nebular contamination in the observed 
spectra of this star, its helium mass fraction can be determined for it is the 
complement of the sum of all other elemental mass fractions considered in the 
calculations), and $X_{O}=0.10 \pm 0.06$. Since no nitrogen line is available in 
the FUSE range, we merely kept the nitrogen mass fraction adopted in our grid 
models.     

\citet{1997IAUS..180..114K} analysed optical and IUE spectra of a series of 
CSPNe, 
among them the central stars of NGC 2867, NGC 5189, and Pb 6 using a non-LTE 
radiative transfer code  
that considers homogeneous winds and He, C, O, and N opacities. They found, for 
NGC 2867, $v_{\infty}=1800$ km s$^{-1}$, $T_{\ast}=141000$ K, and $R_{t}=4.0$ 
R$_{\odot}$. The elemental mass fractions derived by them are $X_{He}=0.66$, 
$X_{C}=0.25$, and $X_{O}=0.09$. Their derived values of $v_{\infty}$, $X_{He}$, 
$X_{C}$, and $X_{O}$ are in agreement within the errors.  

\subsubsection{CSPN Pb 6} \ 

CSPN Pb 6, as well as NGC 5189's central star, displays wider O\,{\sc vi} $\lambda 
\lambda$ 1031.9, 1037.6 $\mathrm{\AA}$ and C\,{\sc iv} $\lambda \lambda$ $1548.2$, 
$1550.8$ $\mathrm{\AA}$ P-Cygni profiles than the other sample objects. We 
determined a terminal velocity of the wind of $v_{\infty}=$2500$\pm 200$ km 
s$^{-1}$ for this star. The strong O\,{\sc vi} $\lambda \lambda$ 1031.9, 1037.6 
$\mathrm{\AA}$ far-UV P-Cygni profile and weak O\,{\sc v} lines observed in CSPN Pb 6 
suggest a temperature of $T_{\ast}=165\pm^{18}_{8}$ kK. The derived transformed 
radius and elemental abundances are: $R_{t}=9.9\pm2.0$ R$_{\odot}$, $X_{C}=0.35 
\pm 0.10$, $X_{He}=0.49 \pm 0.10$, $X_{O}=0.12 \pm 0.06$, $X_{N}=0.03 \pm 
0.01$.

\citet{1997IAUS..180..114K} found for this star: $v_{\infty}=3000$ km s$^{-1}$, 
$T_{\ast}=140000$ K, $R_{t}=4.5$ R$_{\odot}$, $X_{He}=0.617$, $X_{C}=0.24$, 
$X_{O}=0.14$, and $X_{N}=0.003$. Only $X_{O}$ is in agreement with values derived 
in this work within the errors. The stellar temperature derived by us is 
considerably higher and the terminal velocity is much lower. Especially 
discrepant is the nitrogen mass fraction, whose value derived by us is 10 times 
that of the previous analysis.   

\subsubsection{CSPN NGC 5189} \
\label{sec:5189}

As in the case of CSPN Pb 6, we found a terminal velocity of the wind of 
$v_{\infty}=$2500$\pm 250$ km s$^{-1}$ for the central star of NGC 5189. The 
presence of very strong far-UV Ne\,{\sc vii} and O\,{\sc vi} P-Cygni profiles in its FUSE 
spectrum, accompanied by weak O\,{\sc v} lines, led us to derive a stellar temperature 
of $T_{\ast}=165\pm^{18}_{8}$ kK, similarly to Pb 6 and NGC 2867. We derived, 
for the central star of NGC 5189, $R_{t}=9.9\pm2.0$ R$_{\odot}$ (same as for Pb 
6), $X_{C}=0.25 \pm 0.10$ (same as for NGC 2867), $X_{He}=0.58 \pm 0.10$, 
$X_{O}=0.12 \pm 0.06$ (same as for Pb 6), $X_{N}=0.01 \pm 0.01$.

Although we regard the He\,{\sc ii} $\lambda$ 1640.4 $\mathrm{\AA}$ line as a mere upper limit to 
the stellar contribution, due to the evidence of nebular contamination presented 
above, and despite the fact that we derived the parameters of our best-fitting 
model without attempting to precisely match the observed intensity of this line, 
it is, however, reproduced in the final model. In order to maintain the overall 
fit to the other wind lines and at the same time decrease the strength of the He\,{\sc ii} 1640.4 $\mathrm{\AA}$ line, the helium mass fraction would have to be 
severely lowered, at the same time that the value of mass-loss rate would have 
to be decreased as to allow the correspondent increase in carbon and oxygen mass 
fraction that need to occur to assimilate the decrease in the helium mass 
fraction. Mass fractions of neon and nitrogen would also have to be revised up 
to compensate for the lower mass-loss rate. We computed models to explore this 
alternative. They result in considerably worse fits to the O\,{\sc vi} $\lambda 
\lambda$ 1031.9, 1037.6 $\mathrm{\AA}$ doublet, which is now very weak in 
comparison to the observations. This incongruence illustrates the need for higher resolution UV 
spectra of CSPNe. 

The stellar parameters derived by \citet{1997IAUS..180..114K} for NGC 5189 are 
$v_{\infty}=3000$ km/s, $T_{\ast}=135000$ K, $R_{t}=5.0$ R$_{\odot}$, and 
$X_{He}=0.757$, $X_{C}=0.16$, $X_{O}=0.08$, and $X_{N}=0.003$, in mass fractions. 
Only carbon and oxygen mass fractions derived by them agree within the errors 
with those found by us. As in the case of Pb 6, a much higher stellar 
temperature and considerably lower terminal velocity of the wind and helium mass 
fraction were derived by us, besides a 10 times higher nitrogen mass fraction.

\begin{table}
\caption{\small Parameters of our best-fitting models for the sample objects and 
distance dependent parameters. $X_{He}$, $X_{C}$, $X_{N}$, $X_{O}$, and $X_{Ne}$ 
are the 
helium, carbon, nitrogen, oxygen, and neon mass fractions, respectively. For the 
discussion on the derived parameters and errors, see text. Results by other 
authors are also shown. }
\scriptsize
\tabcolsep=0.09cm
\begin{tabular}{ccccccccccccc}
\hline 
Object                               & Pb 6  &         & NGC 6905 &       &      
   & NGC 5189 &         & Sand 3 &              &         & NGC 2867             
 & \\
\hline                   
                                     &t.w.   & KH1997a & t.w.     & M2007 & 
KH1997a & t.w.     & KH1997a & t.w.   & KH1997b      & BH1982  & t.w.            
      & KH1997a\\
\hline
$T_{\ast}$ [kK]                      & 165   & 140     &  150     & 149.6 & 141  
   & 165      &  135    & 150    & 140	         & 200     & 165                 
  & 141  \\
$R_{t}$ [R$_{\odot}$]                & 9.9   & 4.5     &  10.7    & 10.5  & 3.4  
   &  9.9     &  5.0    & 9.4    & 3.0	         &         & 8.5                 
  & 4.0  \\
$v_{\infty}$ [km/s]                  & 2500  & 3000    &  2000    & 1890  & 1800 
   & 2500     &  3000   & 2200   & 2200         & 2700    & 2000                 
 & 1800 \\
$X_{He}$                             & 0.49  & 0.617   &  0.45    & 0.49  & 0.60 
   & 0.58     &  0.757  & 0.28   & 0.615        & 0.38    & 0.60                 
 & 0.66  \\
$X_{C}$                              & 0.35  & 0.24    &  0.45    & 0.40  & 0.25 
   & 0.25     &  0.16   & 0.55   & 0.26         & 0.54    & 0.25                 
 & 0.25  \\
$X_{O}$                              & 0.12  & 0.14    &  0.08    & 0.10  & 0.15 
   & 0.12     &  0.08   & 0.08   & 0.12         & 0.08    & 0.10                 
 & 0.09  \\
$X_{N}$                              & 0.03  & 0.003   &  0.00011 &$<$0.001& 0	 
   & 0.01     &  0.003  & 0.07   & 0.005        &         & 
0.01\tablenotemark{b} & 0  \\
$X_{Ne}$                             & 0.01  &         &  0.02    &       &	 
   & 0.04     &         & 0.02\tablenotemark{b}&&         & 0.04                 
 &   \\                                                  
$\xi_{max}$ [km/s]                   & 200   &         & 150      &       &      
   &  200     &         &  150   &              &         & 200                  
 &   \\
d\tablenotemark{c} [kpc]             &  4.38 &         & 1.70     &       &      
   & 0.55     &         & 0.80\tablenotemark{a}&&         & 1.84                 
 &   \\
\hline
distance dependent &&&&&&&&&&&& \\
parameters &&&&&&&&&&&& \\ 
R$_{\ast}$ [R$_{\odot}$]             &  0.06 &         & 0.09     &       &      
   & 0.03     &         & 0.12\tablenotemark{a} & &        & 0.05                
  &\\
$\log L/L_{\odot}$                   &  3.43 &         & 3.56     &       &      
   & 2.73     &         & 3.84\tablenotemark{a} & &        &3.15                 
  & \\
$log \dot{M}$ [M$_{\odot}$yr$^{-1}$] & -7.29 &         & -7.21    &       &      
   & -7.81    &         & -6.89\tablenotemark{a}& &        & -7.50               
  & \\
\hline
\label{tab:bestfitmodels}
\end{tabular}
\small{t.w.=this work; KH1997a=\citet{1997IAUS..180..114K}; 
KH1997b=\citet{1997A&A...320...91K}; M2007=\citet{2007ApJ...654.1068M}, 
BH1982=\citet{Barlow1982}}
\vspace{-0.2in}
\tablenotetext{a}{Values obtained assuming $\log L/L_{\odot}$=3.84.}
\tablenotetext{b}{Not constrained. Value assumed on the grid models was kept.}
\tablenotetext{c}{Distances from Table \ref{objects}.}
\end{table}
\vspace{0.1in}

\subsection{The problematic near-UV region}

In the studied objects, the spectral region between 1700 and 3200 $\mathrm{\AA}$ 
is problematic. 
The available spectra have worse resolution in this region, which is full of 
structures that make the 
determination of the stellar continuum uncertain. In the region between 1700 and 
2400 $\mathrm{\AA}$, observed features both in absorption and 
in emission are not reproduced by our best-fitting models or grid models of any 
temperature and mass-loss rate. The region between 2150 and 2300 $\mathrm{\AA}$ 
is particularly challenging. 
There is, in this region, a strong feature observed in the central stars of NGC 
6905, NGC 5189, Pb 6, and Sand 3, showing an emission 
component on the red side and an absorption one on the blue side. This feature 
reminds us of a P-Cygni profile in the 
observed spectra of Pb 6 and of NGC 6905 central stars, 
but in the noisier and low resolution spectra available for NGC 5189 and Sand 3, 
the blue side resembles a simple absorption. Our models are able to reproduce 
the red side of these features, in the case of NGC 6905 and of NGC 5189, but not 
the blue absorptions that are located very close to the broad feature of the 
interstellar extinction curve at 2175 $\mathrm{\AA}$, which is related to 
graphite grains \citep{Draine1993}. 
Another difficulty in modelling these stars consists of weak Ne\,{\sc v} and Ne\,{\sc vi} 
lines that are predicted throughout the near-UV by the models, 
but are not observed. While we checked CMFGEN's neon atomic data files for 
consistency with the Atomic Line List 
(\url{http://www.pa.uky.edu/~peter/atomic/}) and found no discrepancy, there may be uncertainties in the atomic data. The neon 
atomic data are primarily taken from the Opacity Project 
\citep{1987JPhB...20.6363S,1994MNRAS.266..805S} and the Atomic Spectra Database 
at the NIST Physical Measurement Laboratory. 

\subsection{Argon and Iron}

\begin{figure}
\begin{center}
\includegraphics[scale=0.42]{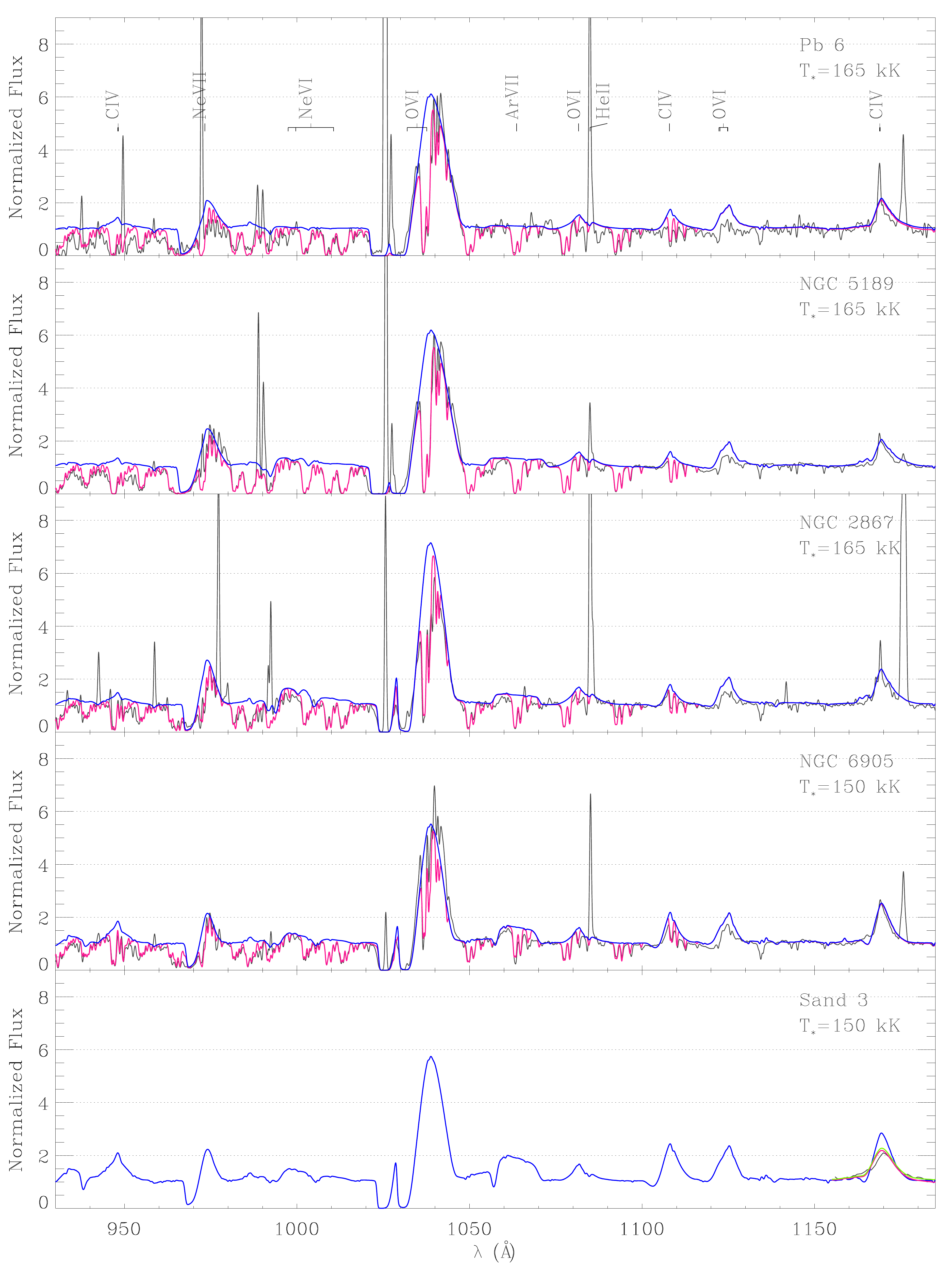}
\caption{FUSE spectra (black line) of CSPNe Pb 6, NGC 5189, NGC 2867, and NGC 
6905 shown together with our best-fitting models, with (pink) and without (blue) 
the effects of interstellar absorption due to molecular and atomic hydrogen. The 
observed and synthetic spectra were degraded to a 0.5 $\mathrm{\AA}$ resolution 
for clarity. No FUSE is spectrum is available for Sand 3. We, however, still 
show the synthetic spectra prediction for this wavelength interval with a 0.5 
$\mathrm{\AA}$ resolution for comparison with results for the other sample 
objects. HST/STIS spectra of this star comprehend the C\,{\sc iv} $\lambda\lambda$ 
1168.8, 1169.0 $\mathrm{\AA}$ line, which is shown here in black. In pink and 
green we show our best-fitting model for Sand 3, with and without Ly\,$\alpha$ 
absorption, convolved with the appropriate HST/STIS LSF. The identifications of 
stellar lines shown in this and following figures correspond to the strongest 
lines predicted by the models at each spectral region.}
\label{allfits_FUV}
\end{center} 
\end{figure}

\begin{figure}
\begin{center}
\includegraphics[scale=0.435]{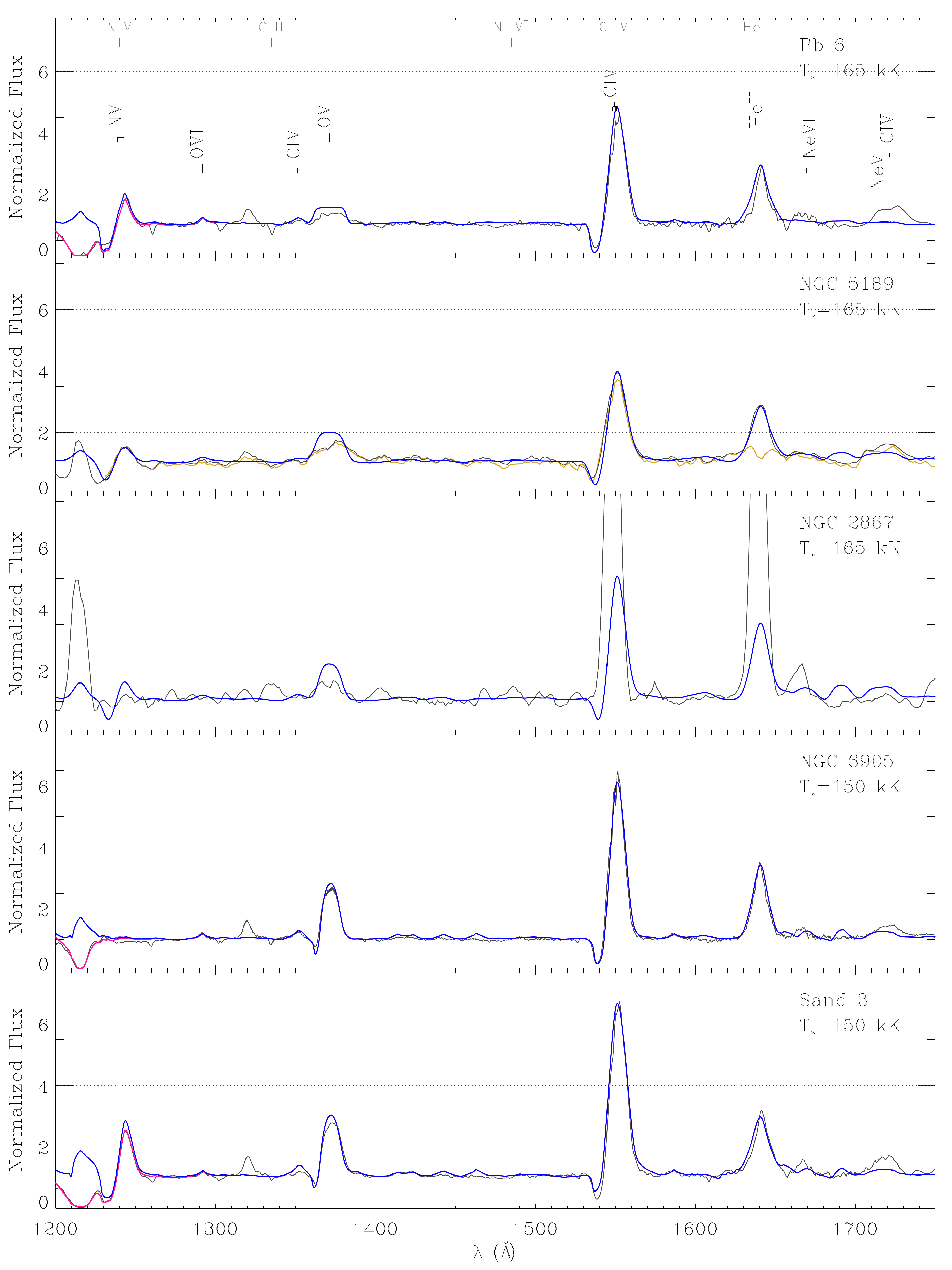}
\caption{Observed spectra of the sample objects (black) in the region between 
1200 and 1750 $\mathrm{\AA}$ shown with our best-fitting models, with (pink) and 
without (blue) ISM absorption, degraded to match the resolution of the 
observations (see table \ref{spectra}). For CSPN NGC 5189 we also show the 
subtracted spectrum from Fig. \ref{subtraction}, in yellow. The grey labels on 
the top part of the figure identify the position of commonly present nebular 
lines.}
\label{allfits_UV}
\end{center} 
\end{figure}

\begin{figure}
\begin{center}
\includegraphics[scale=0.435]{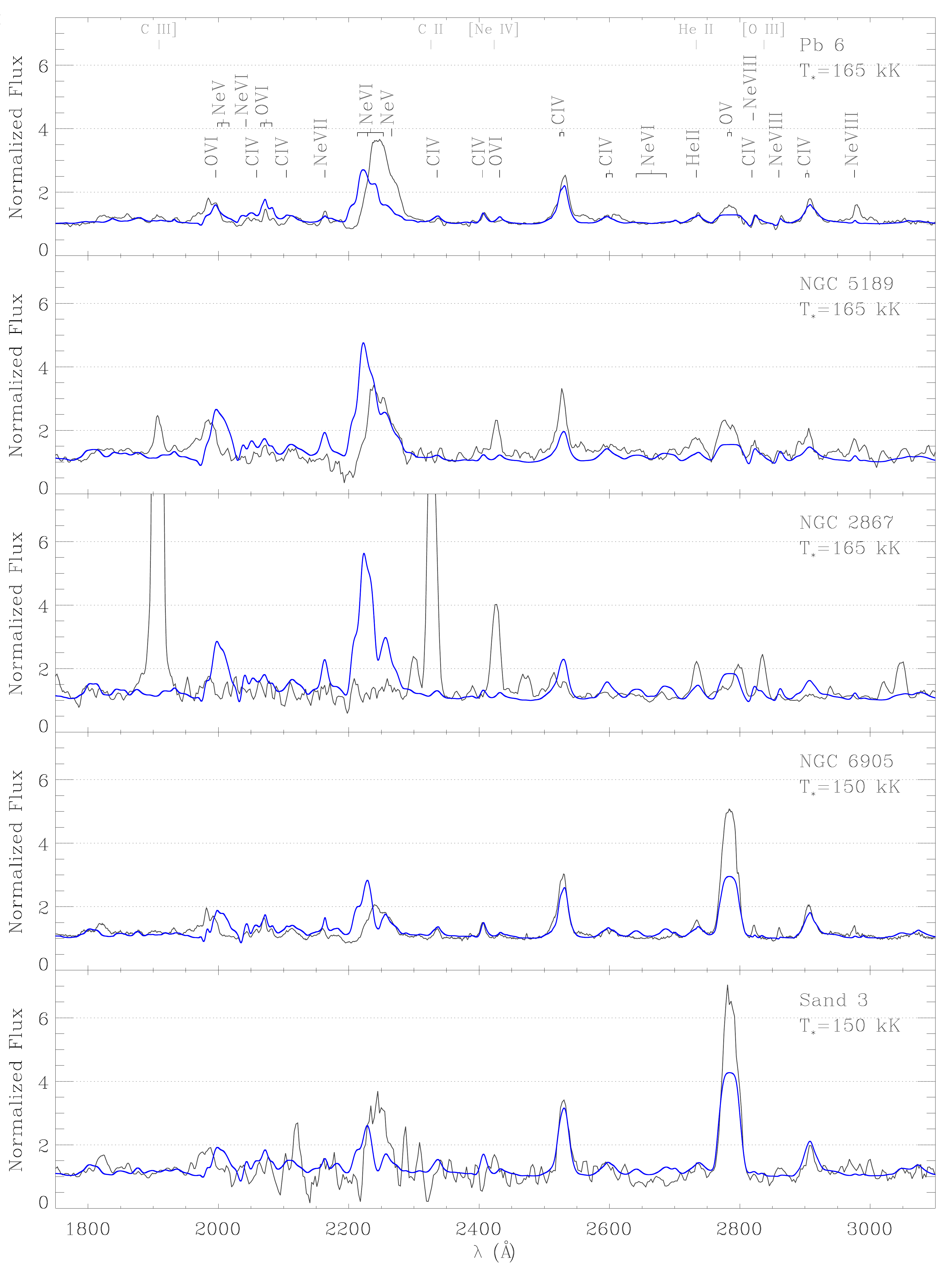}
\caption{Near-UV observed spectra (black) together with our best-fitting models 
to the sample objects (blue).}
\label{allfits_NUV}
\end{center} 
\end{figure}

\begin{figure}
\centering
\includegraphics[angle=0,scale=0.5]{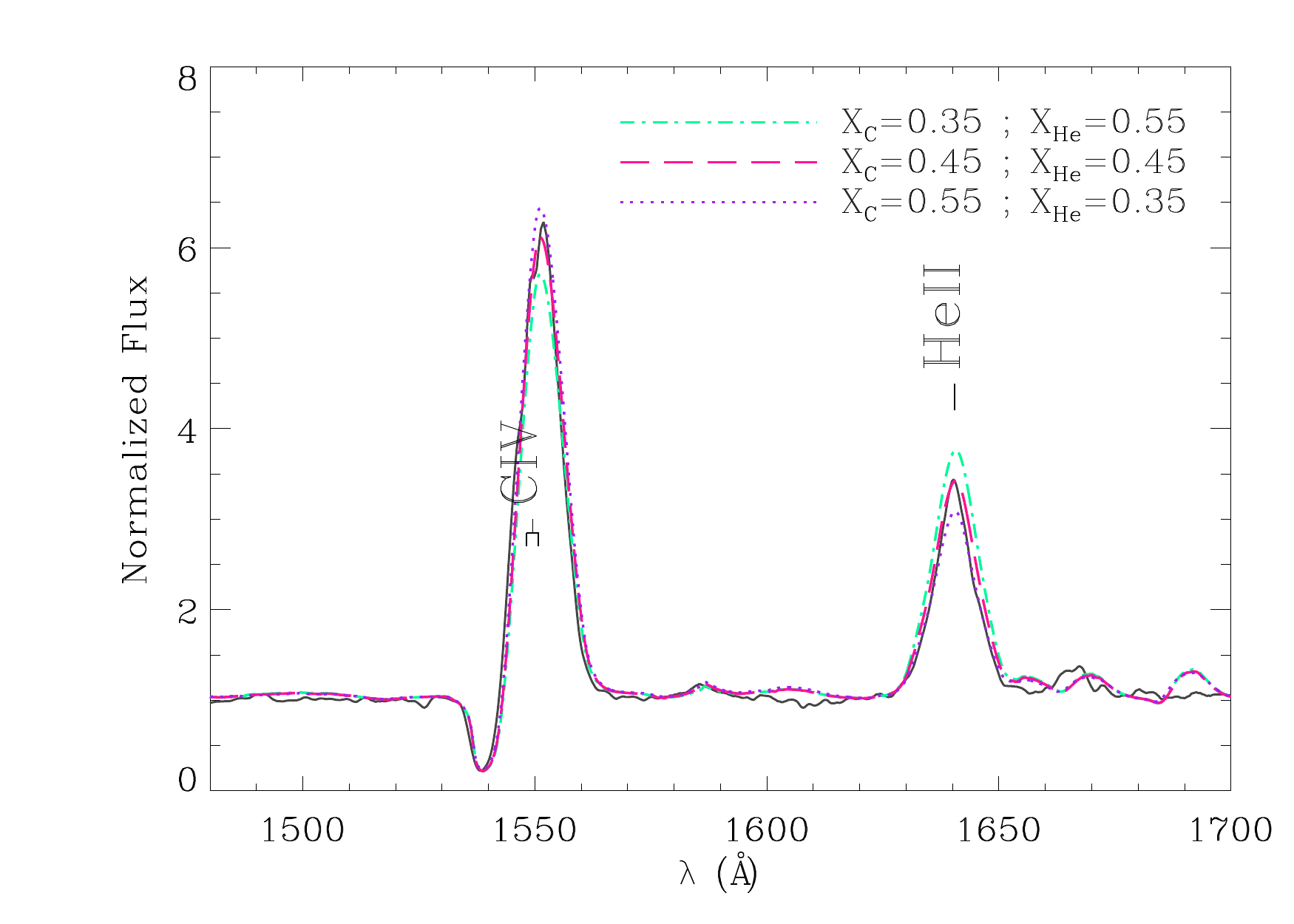}
\caption{Observed spectrum of the central star of NGC 6905 (black continuous 
line) and models with $T_{\ast}=150$ kK,
$R_{t}=10.7R_{\odot}$, $v_{\infty}=2000$ km s$^{-1}$ and different carbon and 
helium abundances.} \label{carbonuncertainty}
\end{figure}

\begin{figure}
\centering
\includegraphics[angle=0,scale=0.5]{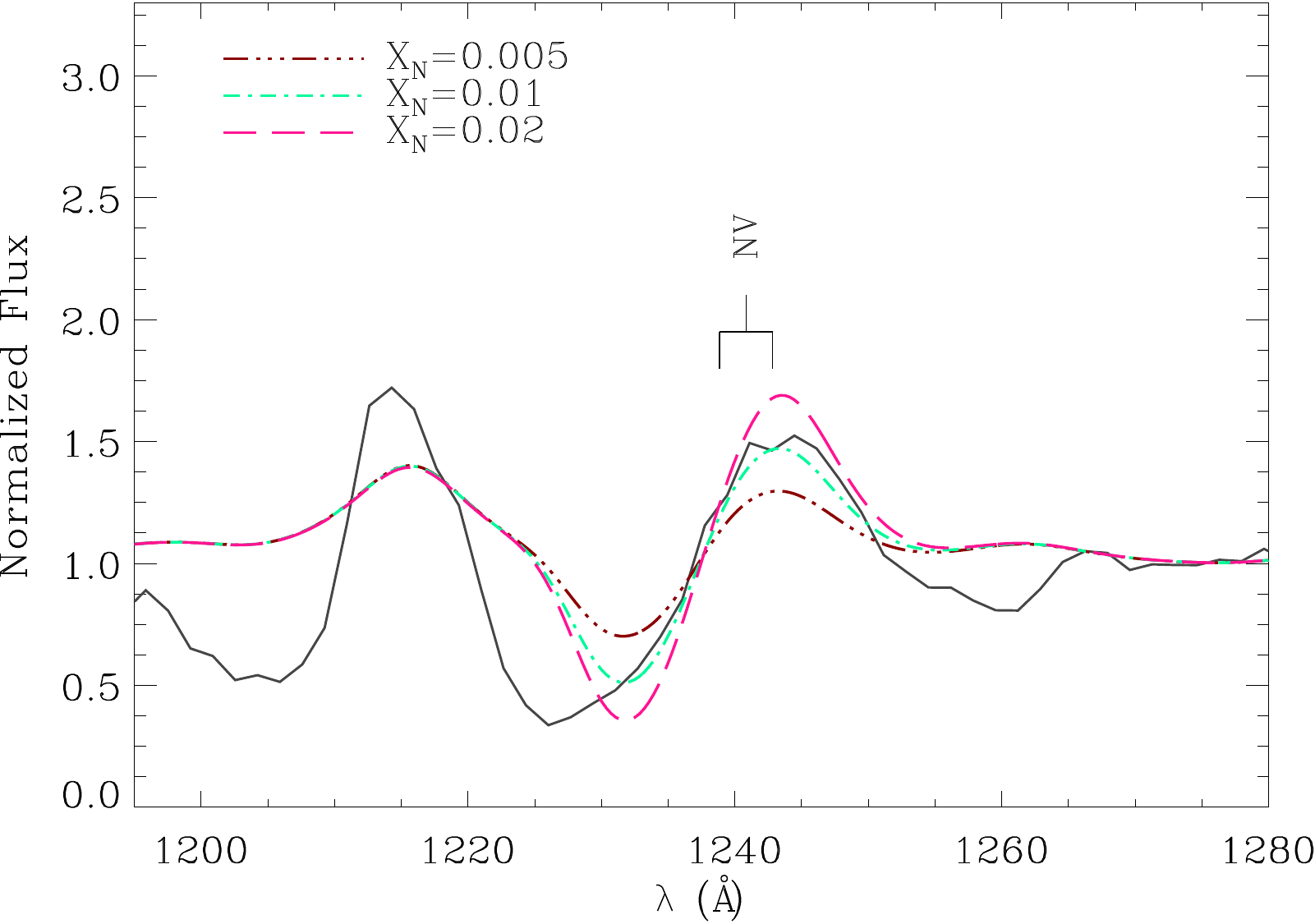}
\caption{Observed spectra of the central star of NGC 5189 (black continuous 
line) and models with $T_{\ast}=165$ kK,
$R_{t}=9.9R_{\odot}$, $v_{\infty}=2500$ km s$^{-1}$ and different nitrogen 
abundances.} \label{nitrogenuncertainty}
\end{figure}

In \citet{kelleretal2011PORT}, we found that a model with 10 times the argon 
solar abundance produces an Ar\,{\sc vii} $\lambda$ 1063.55 $\mathrm{\AA}$ line which 
approximately reproduces the observed line profile in that spectral region of 
CSPN NGC 6905. In Fig. \ref{allfits_FUV}, the best-fitting models for NGC 5189 
and NGC 2867 also have the same argon overabundance. However, the shapes of 
the observed and synthetic profiles disagree, challenging the identification of 
this line.

In all best-fitting models, we assumed a strong iron underabundance of one 
hundredth of the solar value, since higher values would entail the appearance of a number of non-observed Fe\,{\sc x} lines in the synthetic spectra 
\citep{kelleretal2011PORT}. These lines are not, however, observed in PG1159 
stars for which solar iron abundances were found by \citet{Werner2010,Werner2011a} and there may be strong uncertainties in their computed 
wavelengths.    

\subsection{Reddening}

\begin{figure}
\begin{center}                                                      
\includegraphics[width=0.9\linewidth,height=0.185\textheight]{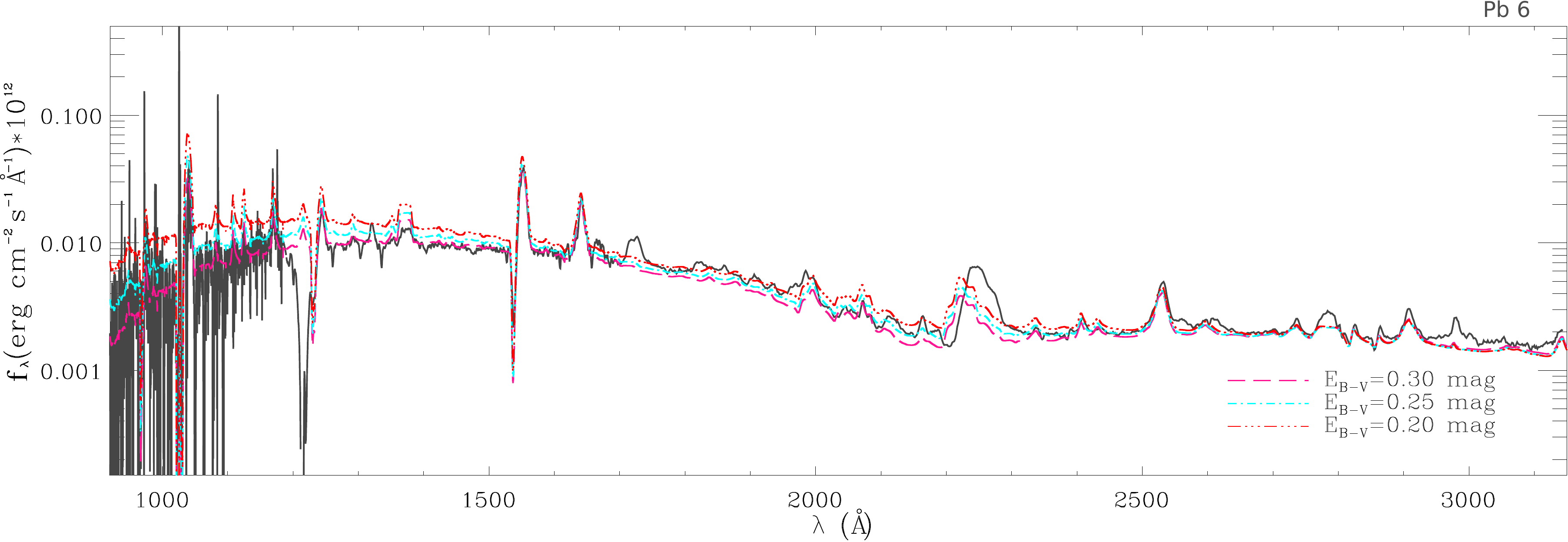}
\includegraphics[width=0.9\linewidth,height=0.185\textheight]{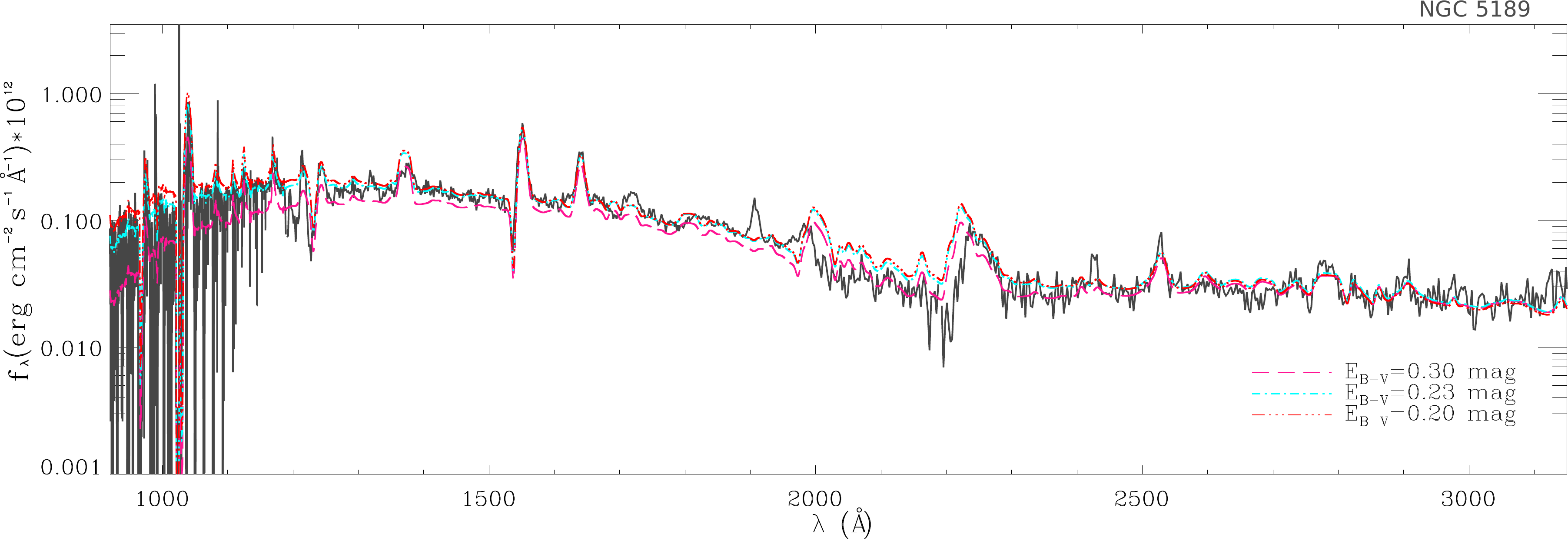}
\includegraphics[width=0.9\linewidth,height=0.185\textheight]{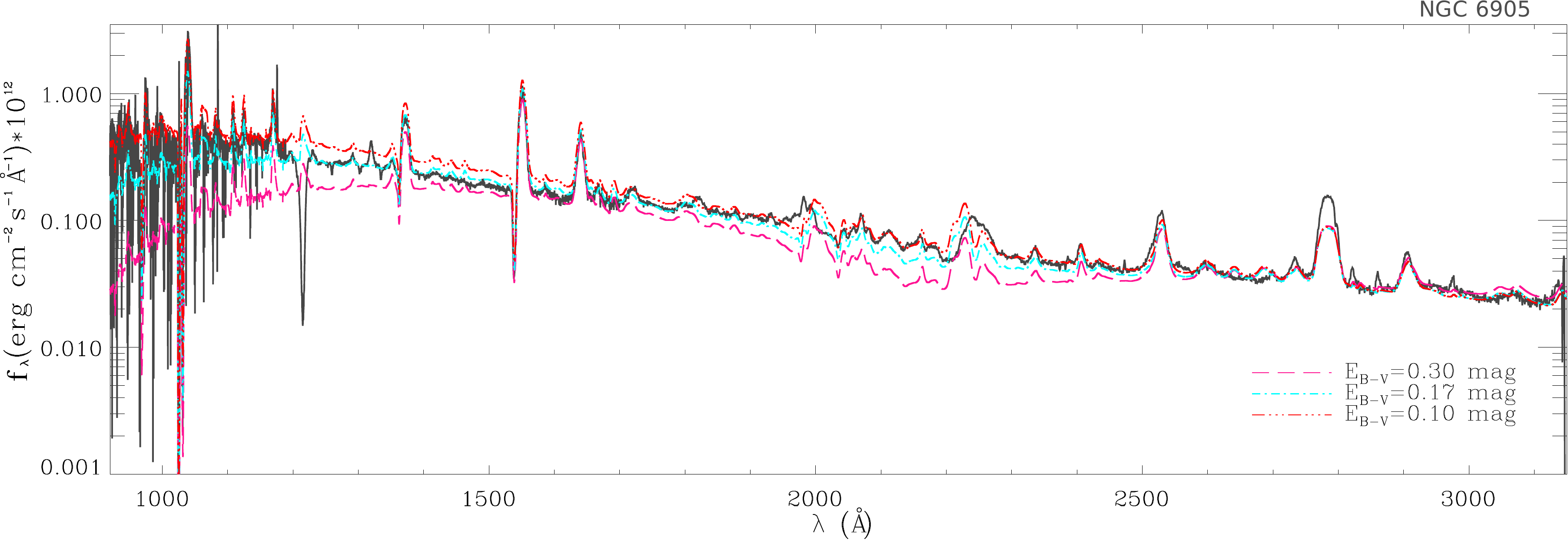}
\includegraphics[width=0.9\linewidth,height=0.185\textheight]{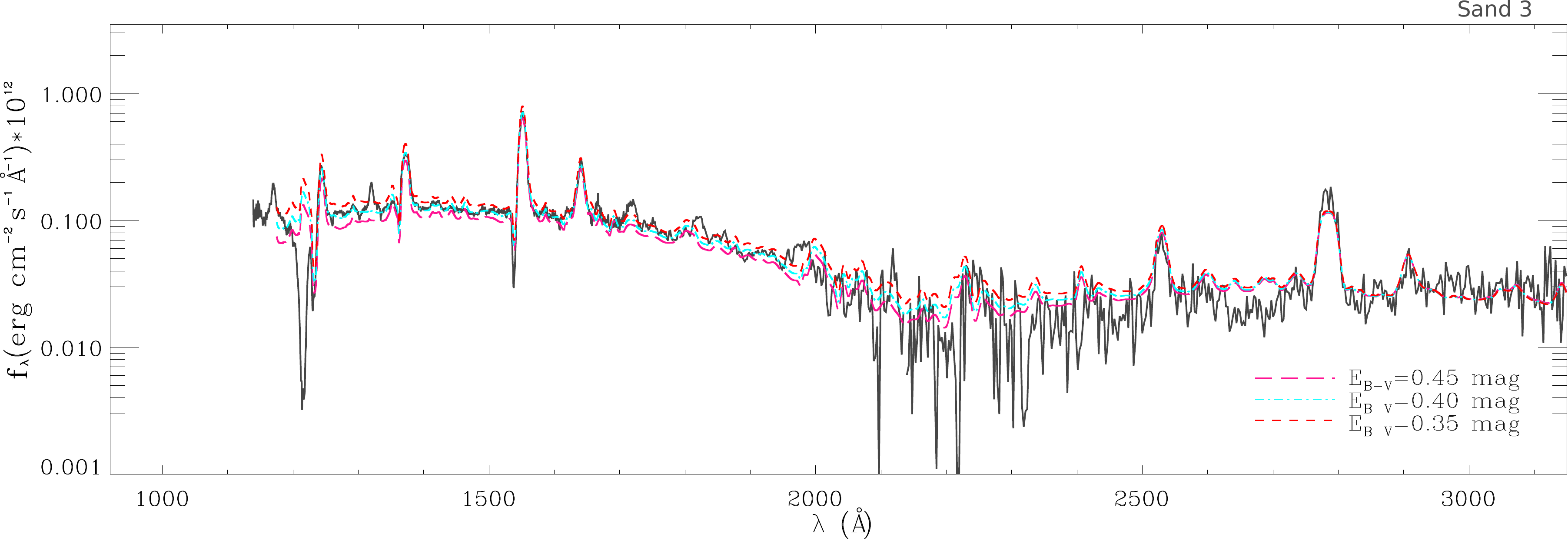}
\caption{The observed spectra of the sample objects (continuous black lines), 
except for NGC 2867, 
are shown together with our best-fitting models reddened by different values of 
the colour excess. 
We used the \citet{Cardelli1989} extinction curves, with $R_{V}=3.1$.}
\label{Reddening}
\end{center} 
\end{figure}

\begin{figure}
\begin{center}                                                      
\includegraphics[width=0.9\linewidth,height=0.185\textheight]{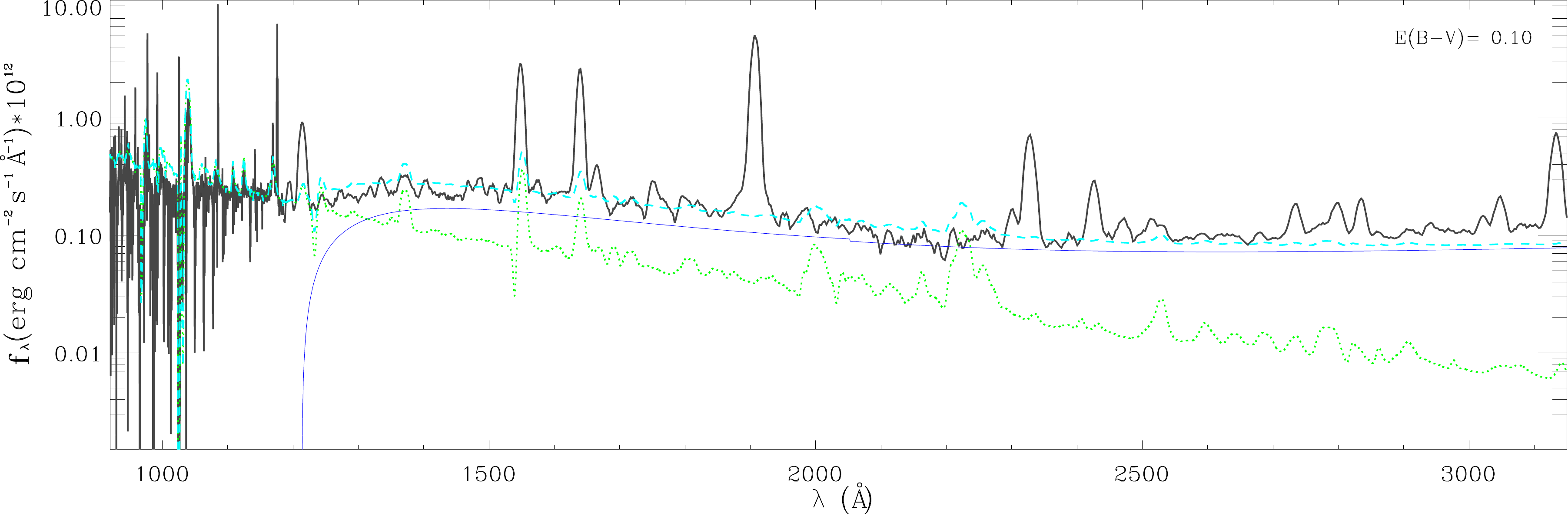}
\includegraphics[width=0.9\linewidth,height=0.185\textheight]{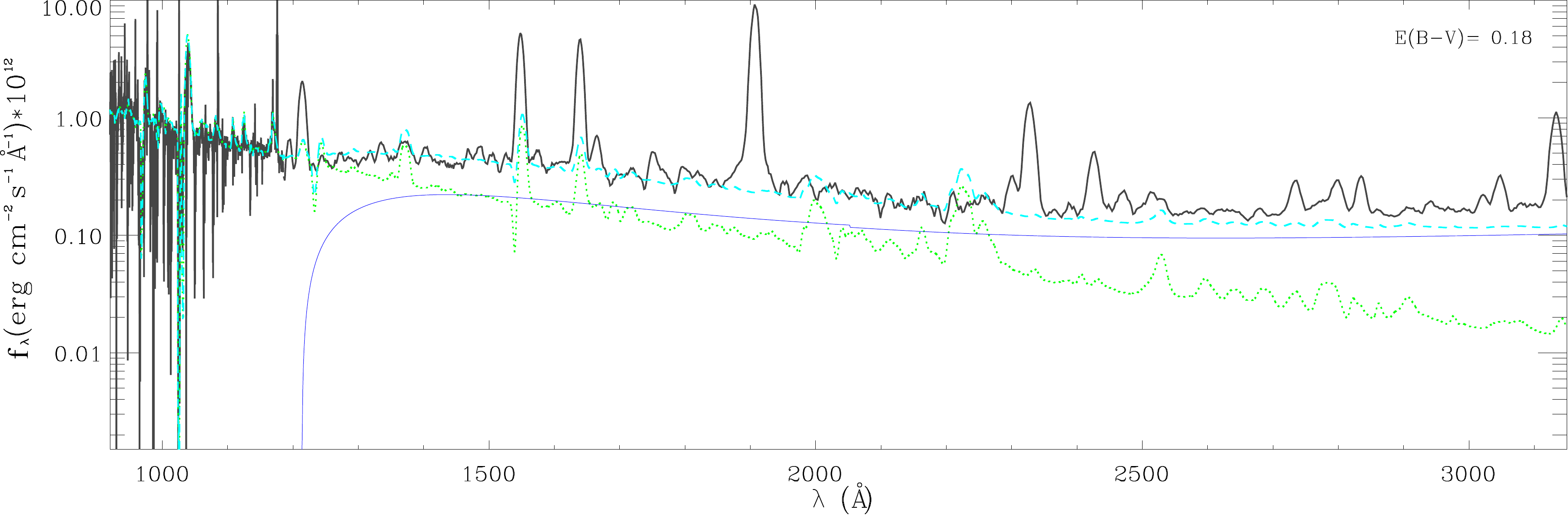}
\includegraphics[width=0.9\linewidth,height=0.185\textheight]{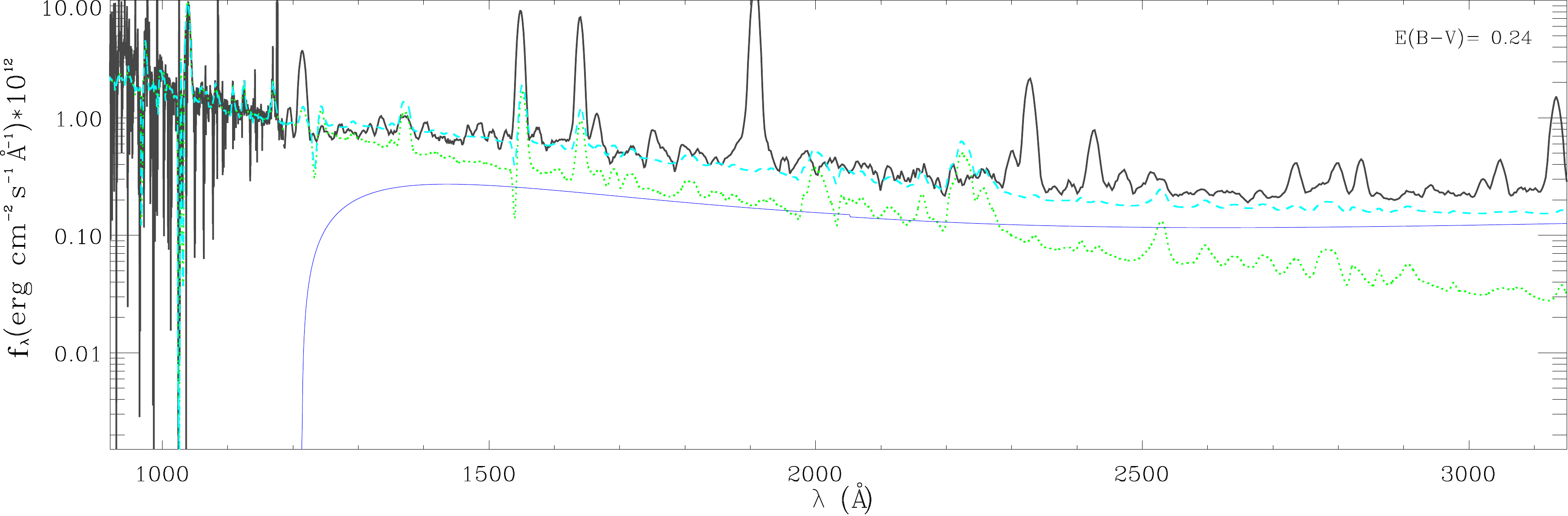}
\caption{The observed spectra of NGC 2867 (continuous black lines) dereddened 
adopting different values of the colour excess is shown together with our 
best-fitting model for the stellar spectrum (green dotted line), a model for the 
nebular continuum (blue continuous line) and the sum of both models (cyan dashed 
line). 
We used the \citet{Cardelli1989} extinction curves, with $R_{V}=3.1$.}
\label{NGC2867Reddening}
\end{center} 
\end{figure}

Having constrained the main stellar parameters through the analysis of the 
strength of spectral lines, we now compare the slope of the observed 
spectral continuum of each sample object to that of our best-fitting models 
reddened by different values of colour excess as shown in Fig. \ref{Reddening}. 
We adopted \citet{Cardelli1989} extinction curves with $R_{V}=3.1$. The results 
are listed in Table \ref{tab:is}. In the case of NGC 2867, which is shown 
separately in Fig. \ref{NGC2867Reddening}, we had to take into consideration the 
contribution from the nebular continuum, which heavily affects 
wavelengths longer than Ly\,$\alpha$. We estimated the nebular continuum 
emission using the code described in \citet{1997ApJ...480..290B}, which accounts 
for two-photon, H and He recombination, and free-free emission processes. 
Nebular parameters were taken from the literature 
\citep{2002ApJS..138..285M,2002ApJS..138..279M} and they include the electron 
density, electron temperature, observed H$\beta$ flux, helium-to-hydrogen ratio, 
ratio of doubly to singly ionized helium and the extinction quantity ($c$). 
We found $E(B-V)=0.18$, which corresponds to the value of 
$c$ found in the literature, to allow for an overall good match between observed 
continuum of NGC 2867 and the sum of model stellar and nebular emissions.

\begin{figure}
\begin{center}
\includegraphics[scale=0.435]{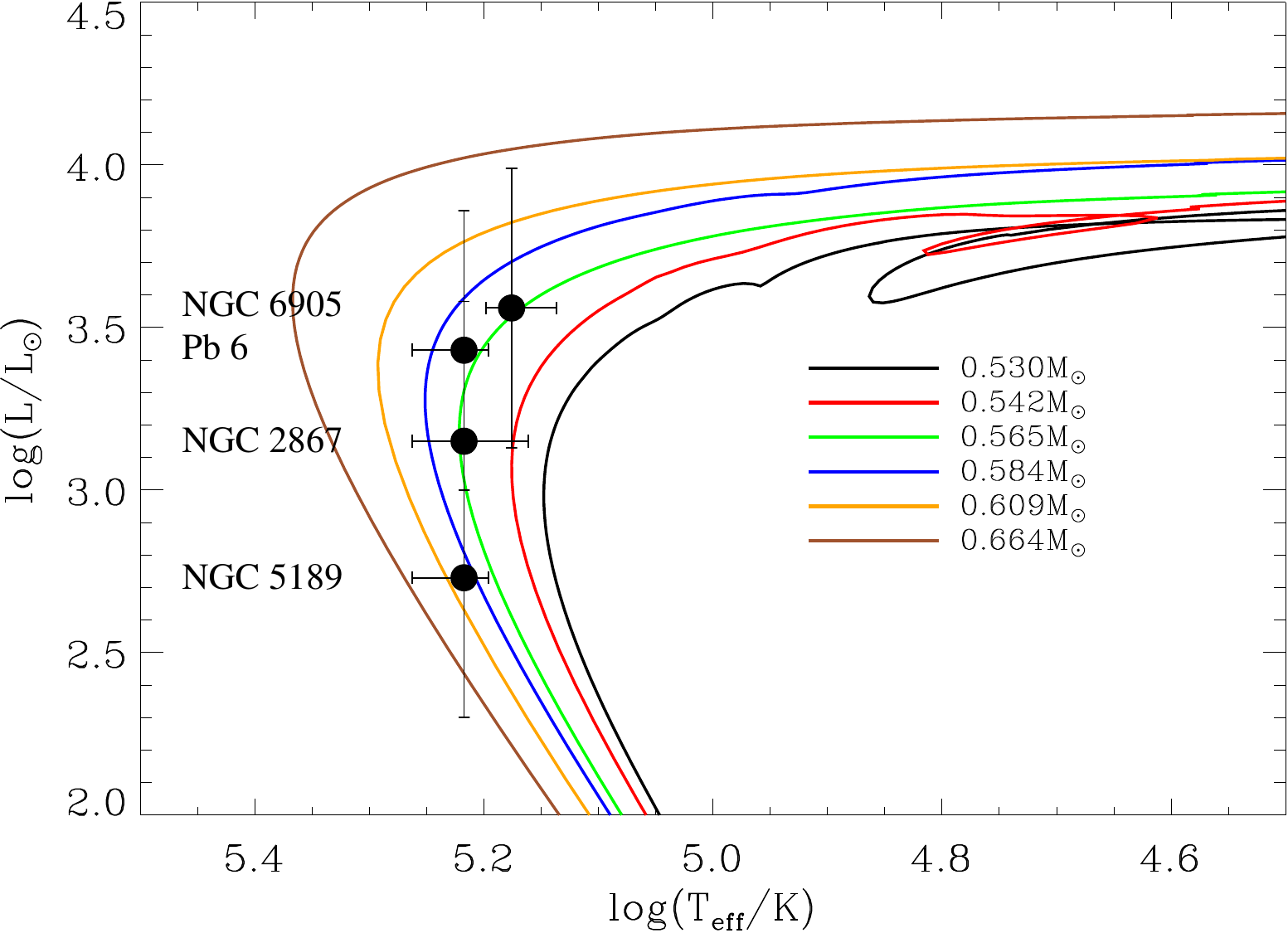}
\caption{HR diagram with evolutionary tracks from \citet{2006A&A...454..845M}. 
The open circles correspond to the sample objects. Sand 3 was omitted from it, 
because, to the best of our knowledge, there is no measurement of its distance 
in the literature.}
\label{TxL_GRID}
\end{center} 
\end{figure}

\subsection{Discussion}
\label{subsec:abund}

Fig. \ref{TxL_GRID} shows the evolutionary tracks of \citet{2006A&A...454..845M} 
in the $\log(T_{eff}) \times \log(L)$ diagram and the position of our sample 
objects. Luminosities are based on distances taken from the literature and are 
shown in Table \ref{tab:bestfitmodels}. Sand 3 was omitted from the figure, 
because we were unable to find measurements of its distance in the literature. 
The sample objects intercept the evolutionary tracks at 0.565 and 0.584 solar 
masses. The distance values adopted from the literature place the central stars 
of NGC 6905 and Pb 6 on the verge of entering the WD cooling track, while they 
put the central stars of NGC 5189 and NGC 2867 already within it, in a region 
where PG1159 stars are expected. [WCE]-type CSPNe are, however, expected to 
populate the region of constant luminosity of the evolutionary tracks in the 
HRD, which, at the temperatures derived here, would require higher luminosities 
and thus, greater distances.

\begin{figure}
\begin{center}
\includegraphics[scale=0.435]{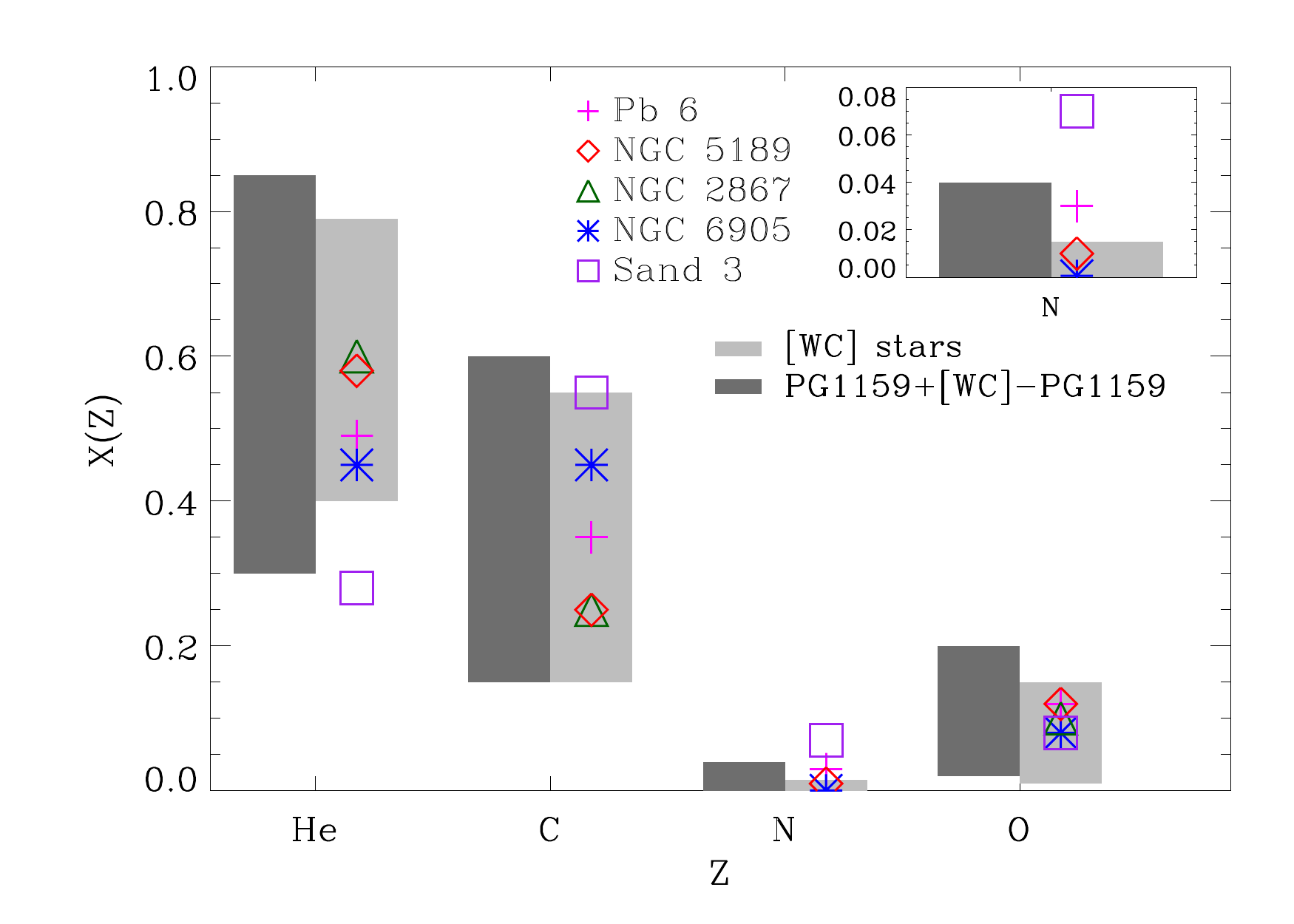}
\caption{The abundance intervals of PG1159, [WC]-PG1159 and [WC] stars (as found 
from the literature) are compared to the values determined in this work.}
\label{abunds}
\end{center}
\end{figure}

Among the five objects analysed in this work, Sand 3 is the only one whose 
derived abundances 
seem to place it outside the range of values found in the literature for [WC] 
stars, since it shows, 
according to Table \ref{tab:bestfitmodels}, a quite low helium abundance of 0.28 
in mass fraction (not significantly lower, however, than the lowest value of 
helium mass fraction found for PG1159 objects, which is 0.30). 
The derived He, C, and O abundances, in mass fractions, of our sample objects are 
in the intervals: $X_{He}=0.28-0.60$, $X_{C}=0.25-0.55$, and $X_{O}=0.08-0.12$ 
(see Fig. \ref{abunds}). It is generally believed that [WCL]-type CSPNe 
evolve into [WCE] objects and thus, both classes are expected to present 
somewhat similar carbon to helium ratios. However, previous analyses have shown 
the carbon content in [WCE]-type stars to be lower than what is seen in [WCL] 
objects, with 0.19$<C:He<$0.7 for [WCE] stars 
\citep{1997IAUS..180..114K,1997A&A...320...91K} and $C:He$ typically higher than 
1 for [WCL] stars \citep{1996A&A...312..167L}. The wide range of values found by 
us in this work seems to argue against this notion: we found 
$C:He=$0.42, 0.43, 1.0, 0.71, 1.96 for the central stars of NGC 2867, NGC 5189, 
NGC 6905, Pb 6, and Sand 3, respectively. This is in line with the results from 
\citet{2003IAUS..209..243C} and \citet{2007ApJ...654.1068M} that found no 
systematic discrepancies in the $C:He$ ratios between [WCL] and [WCE] stars.

Values of the nitrogen mass fraction derived in the literature range from 
$< 3 \times 10^{-5}$ \citep[see][and references therein]{2006PASP..118..183W} 
to 0.04 \citep{Leuenhagen1993}. We have derived an upper limit of 
$5.5\times10^{-4}$ for the nitrogen abundance of NGC 6905, which would indicate 
it is the product of a LTP or even an AFTP. The central stars of NGC 5189, Pb 6 
and Sand 3, on the other hand, all show conspicuous N\,{\sc v} $\lambda \lambda$ 
1238.8, 1242.8 $\mathrm{\AA}$ lines that led to derived 
nitrogen mass fractions of 1, 3, and 7 per cent, respectively, pointing to the 
occurrence of VLTP events. 

We have not included hydrogen in our grid models or in the best-fitting models 
presented in this work. However, we calculated exploratory models which indicate 
an upper limit to the hydrogen mass fraction of 20 per cent for our sample 
objects.  

Neon abundances of about 2 per cent have been reported for a number of PG1159 
stars \citep{2004A&A...427..685W}, which is in agreement with stellar evolution 
models that predict a neon mass fraction of about 2 per cent in the intershell 
region. In this work, modelling of the far-UV Ne\,{\sc vii} and Ne\,{\sc vi} features led us 
to infer neon mass fractions between 1 and 4 per cent.    

\section{Conclusions}
\label{sec:conclusions}

We have analysed UV spectra from 5 of the hottest known [WCE]-type CSPNe, which are 
the central stars of NGC 2867, NGC 5189, NGC 6905, Pb 6, and Sand 3. The 
analysis was performed using our 
extensive grid of synthetic spectra, calculated with CMFGEN, which allowed us 
to constrain temperatures, transformed radii, and terminal velocities of the 
wind, in a uniform and systematic way. The 
analysis was then refined by exploring other parameters, such as elemental abundances and microturbulence velocities, besides adding 
further elements and ionic species into the calculations.

We applied the effects of interstellar absorption due to neutral and
molecular hydrogen to our best-fitting
models and determined their column densities and after the stellar parameters 
were constrained based on line diagnostics, we compared the
observed slope of the stellar continua to our best-fitting models reddened using 
different values
of colour excess.

We derived $T_{\ast} = 150000$ K for CSPNe NGC 6905 and Sand 3 and $T_{\ast} = 
165000$ K for the central stars of NGC 2867, NGC 5189, and Pb 6. This result 
corresponds to an upward revision of the temperatures derived for the latter 
three objects, in comparison with previous analyses (for Sand 3 and the 
central star of NGC 6905, the derived temperatures agree with previous results 
within the errors), a fact that has two reasons: the first one is the inclusion 
of several less abundant ions into the calculations, as described above, that 
caused the O\,{\sc v} lines to increase in intensity without altering other line 
diagnostics, allowing us to fit the observed spectra with hotter models. The O\,{\sc v} 
lines are important temperature diagnostics and allowed us to separate between 
models of 150 and 165 kK; The second reason is the modelling of the O\,{\sc vi} $\lambda 
\lambda$ 1031.9, 1037.6 $\mathrm{\AA}$ and Ne\,{\sc vii} $\lambda$ 973 $\mathrm{\AA}$ 
lines in the FUSE range, which had not been modelled in previous analyses of 
these three stars and which are stronger in our grid models of 165 kK. The 
terminal wind velocities of the sample objects were found to range from 2000 to 
2500 km s$^{-1}$. Despite the fact that the five stars studied here present 
somewhat similar spectra, and that
we derived similar temperatures for them, we derived quite different helium, 
carbon
and nitrogen abundances. The values of carbon to helium mass ratio found by us 
for the sample objects span a wide range of values: $0.42<C:He<1.96$. The case 
of Sand 3 seems an intriguing
one. Not only we find a very low helium abundance for this star, but we also 
derive a high 
nitrogen content, when comparing these with values found in the literature for 
[WC] and PG1159
stars. Nitrogen abundances derived by us for the central stars of NGC 5189, Pb 6 
and Sand 3 were found to be higher than previous analyses by factors of 3, 10 
and 14, respectively.

By considering many elements and ionic species neglected in previous analyses, we improved the fits to the observed spectra: we were able to 
reproduce the observed
strengths of the O\,{\sc v} $\lambda$ 1371.3 $\mathrm{\AA}$ lines and improve the fits 
of the 
O\,{\sc v} $\lambda \lambda$ 2781.0, 2787.0 $\mathrm{\AA}$ lines, 
thanks to the consequent increase of the O\,{\sc v} ionization fractions and we were 
also provided with the Ne\,{\sc vii} $\lambda$ 973 $\mathrm{\AA}$ line diagnostic, 
which was modelled for the first time in [WCE] stars. The modelling of the 
far-UV Ne features points towards strong neon overabundances for the central 
stars of NGC 6905, NGC 5189, NGC 2867, and Pb 6, with Ne mass fractions between 
0.01 and 0.04.  
  
The near-UV region remains problematic: observed features both in absorption 
and 
in emission are not reproduced by our best-fitting models or grid models of any 
temperature and mass-loss rate and weak Ne\,{\sc v} and Ne\,{\sc vi} lines are predicted by 
the models throughout the region, but are not observed. 

\vspace{1cm}

\textit{Acknowledgements.} We thank Klaus Werner for providing many useful 
comments. We are also grateful to an anonymous referee for several comments
that helped us improve the present paper. G.R. Keller gratefully acknowledges financial support from the
Brazilian agencies CAPES (which supported her work at the Johns Hopkins 
University, Department of Physics and Astronomy) and FAPESP, 
grants 0370-09-6, 06/58240-3, and 2012/03479-2. This work was based on 
observations made with the Far Ultraviolet Spectroscopic Explorer, 
the Hubble Space Telescope, and the International Ultraviolet Explorer. The 
data presented 
in this paper were obtained from the Mikulski Archive for Space Telescopes 
(MAST). STScI is operated by the Association of Universities for Research in 
Astronomy, Inc., under NASA contract NAS5-26555. Support for MAST for non-HST 
data is provided by the NASA Office of Space Science via grant NNX13AC07G and by 
other grants and contracts. This work has made use of the computing facilities 
of the Laboratory of Astroinformatics (IAG/USP, NAT/Unicsul), whose purchase was 
made possible by the Brazilian agency FAPESP (grant 2009/54006-4) and the 
INCT-A.

\bibliographystyle{grk_pp_mnras}
\bibliography{refs_keller}

\begin{appendices}
\section{Online-only figures}

\begin{figure}
\centering
\includegraphics[scale=0.5]{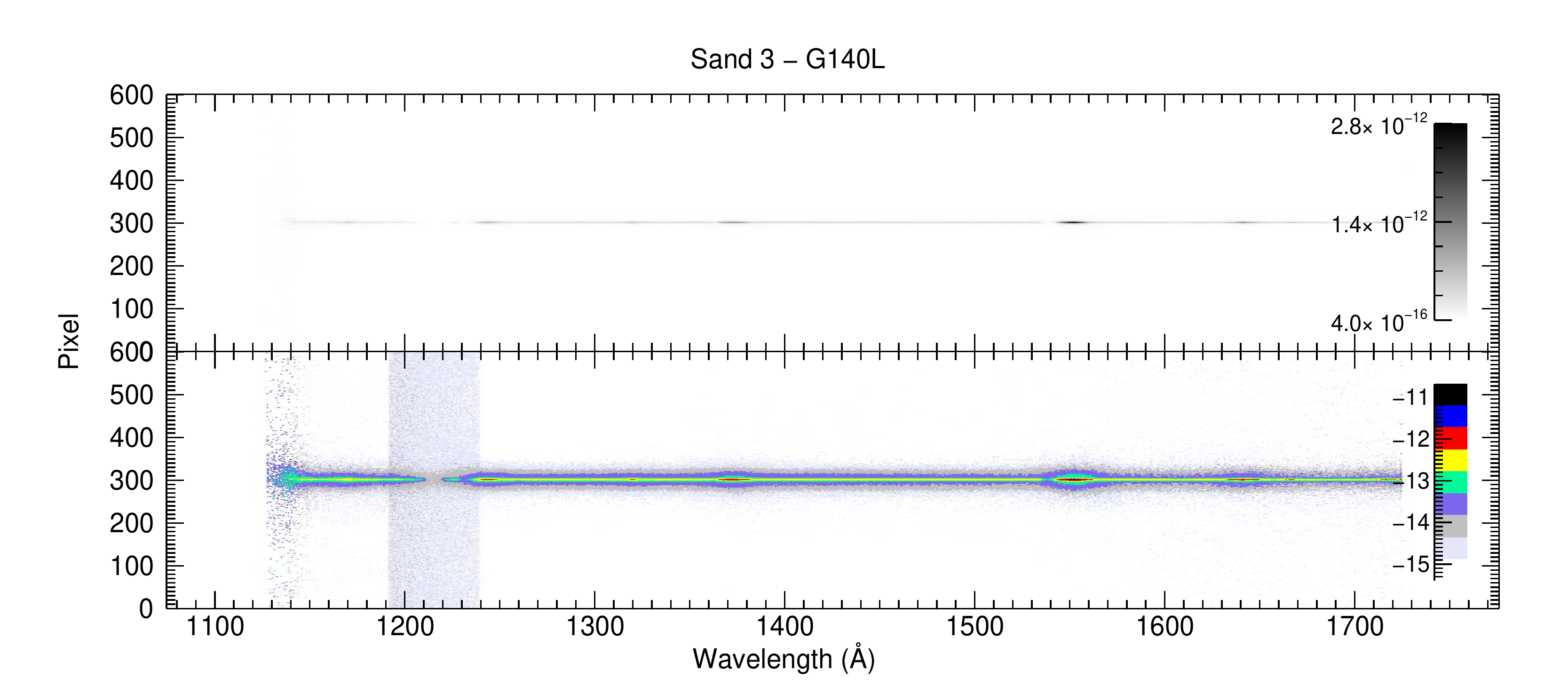}
\includegraphics[scale=0.5]{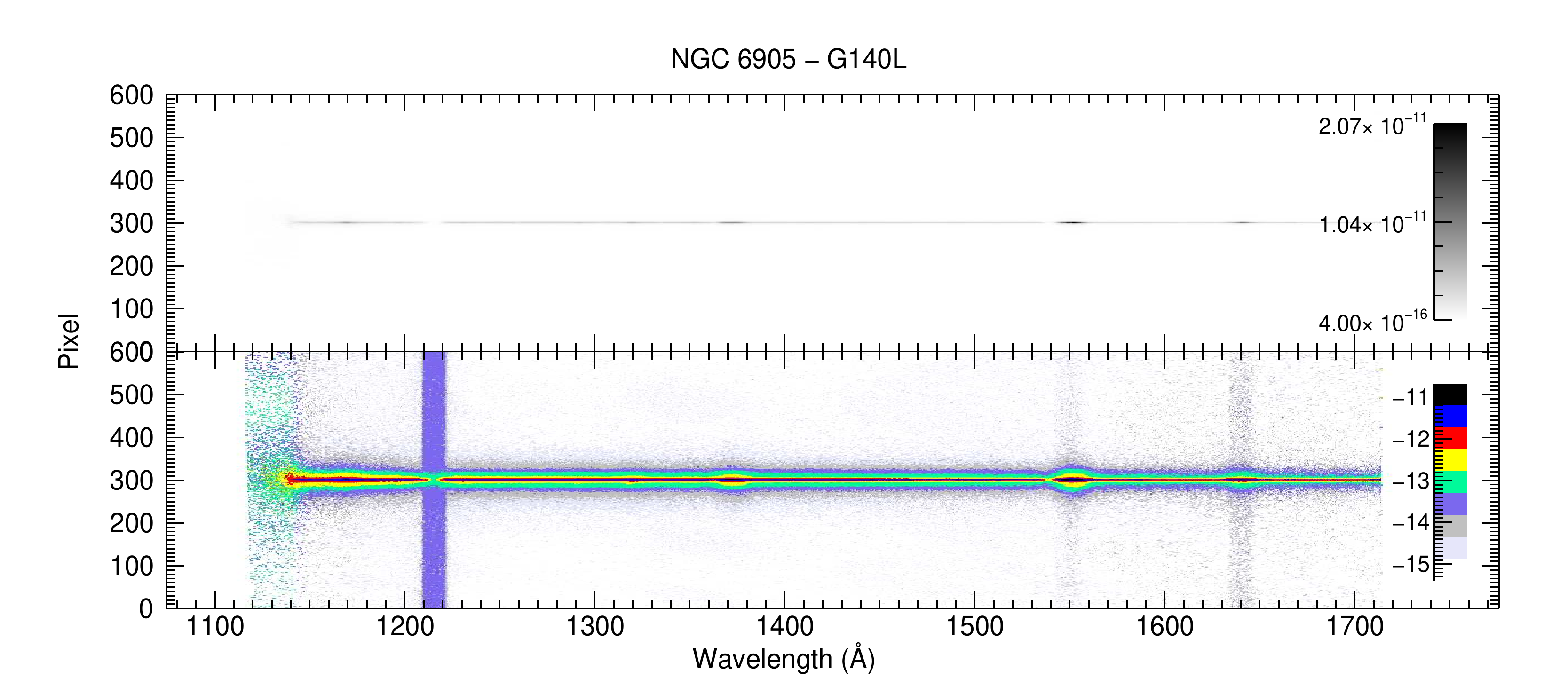}
\includegraphics[scale=0.66]{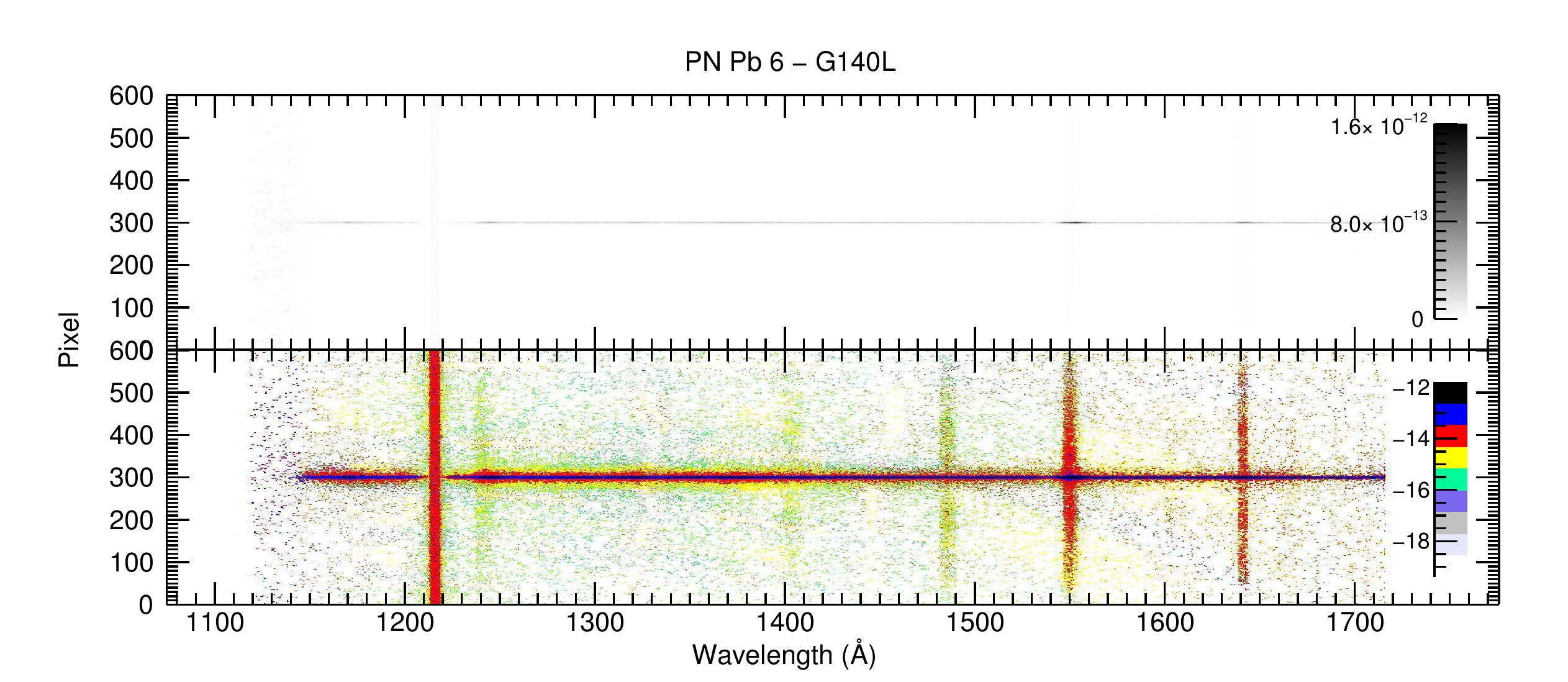}
\caption{HST/STIS 2D Sand 3 G140L (top panel), NGC 6905 G140L (central panel), 
and Pb 6 G140L (bottom panel) spectra. Each spectrum is shown in linear grey 
scale (top) and logarithmic scale (bottom). The spatial direction comprehends 
14.4 arcseconds. The vertical bands around 1216 $\mathrm{\AA}$ are due to 
geocoronal Ly-$\alpha$ emission. The objects were observed using different slit 
widths, as described in Table \ref{spectra}.} \label{2dcolormap}
\end{figure} 

\begin{figure}
\centering
\includegraphics[scale=0.5]{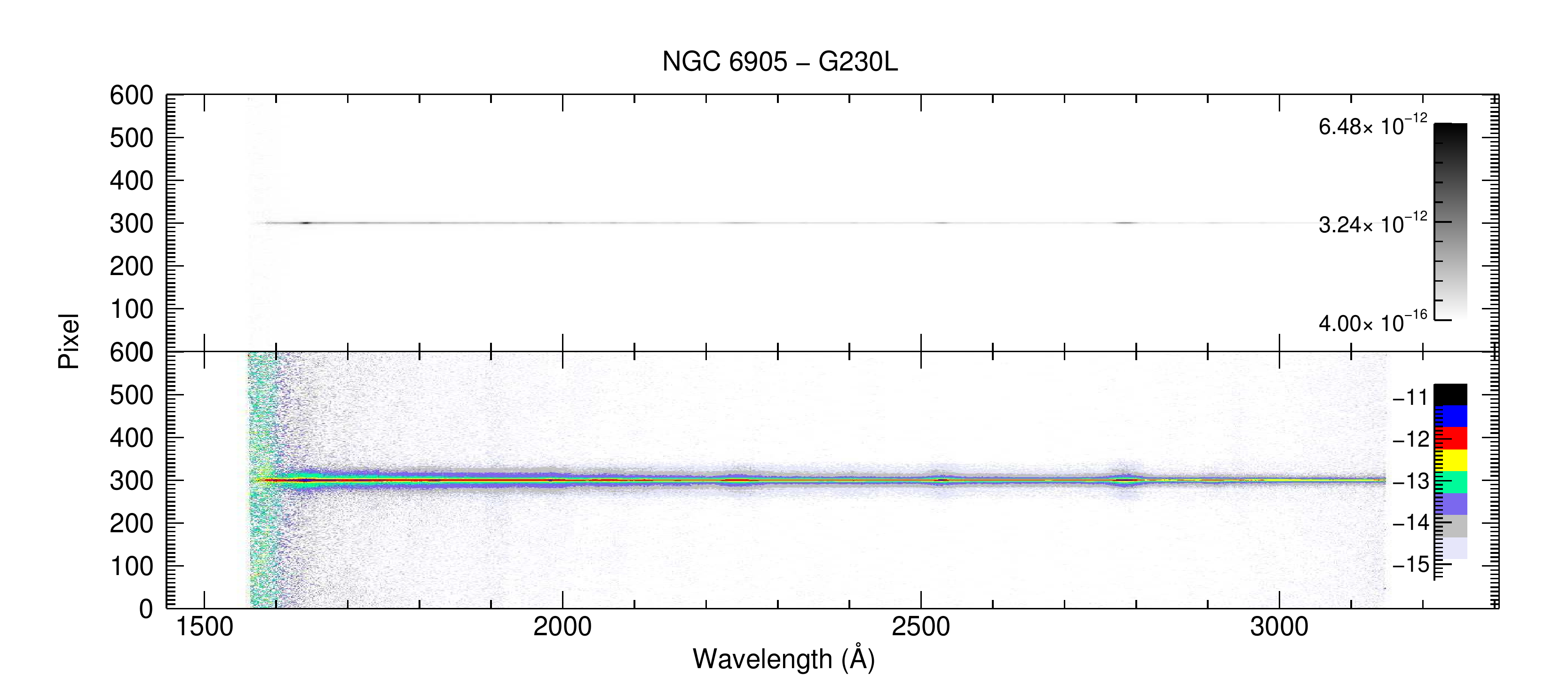}
\includegraphics[scale=0.66]{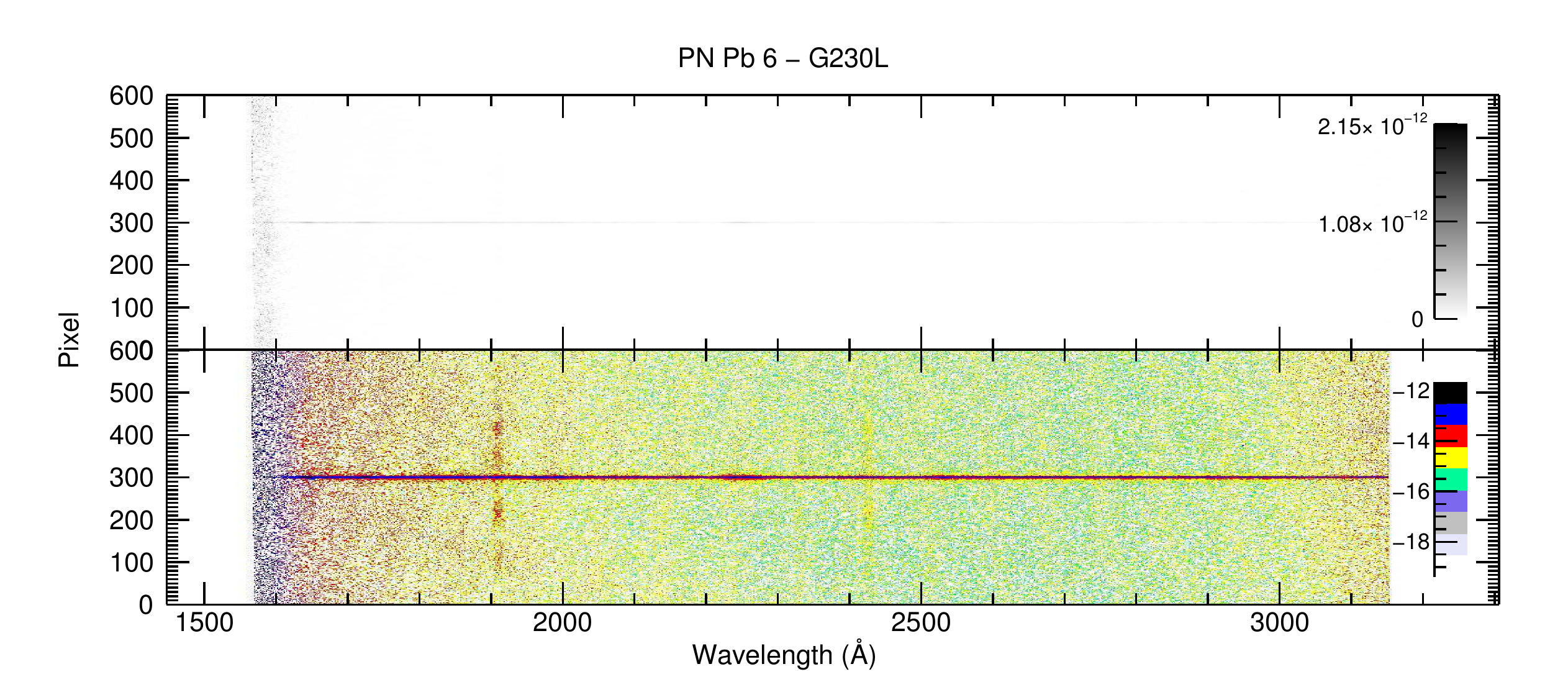}
\caption{HST/STIS 2D NGC 6905 G230L (top panel) and Pb 6 G230L (bottom panel) 
spectra. Each spectrum is shown in linear grey scale (top) and logarithmic scale 
(bottom). The spatial direction extends for 15 arcseconds.} \label{2dcolormap_b}
\end{figure}

\end{appendices}

\label{lastpage}

\end{document}